\documentclass[a4paper]{amsart}

\usepackage{lineno}
\usepackage[hidelinks]{hyperref}

\usepackage{natbib}

\usepackage{amsmath}
\usepackage{amssymb}
\usepackage{amsthm}
\usepackage{hyperref}
\usepackage{subfig}
\usepackage{graphicx}
\usepackage{graphbox} 
\usepackage[left=3cm, right=3cm, bottom=3cm, top=3cm]{geometry}

\usepackage{float}
\usepackage{xcolor}
\hypersetup{
    colorlinks,
    linkcolor={red!50!black},
    citecolor={blue!50!black},
    urlcolor={blue!80!black}
}
\usepackage[foot]{amsaddr}

\theoremstyle{plain} 

\DeclareMathOperator{\sgn}{sgn}

\usepackage{mathtools}
\usepackage{pgfplots}
\pgfplotsset{/pgf/number format/use comma,compat=newest}

\renewcommand\epsilon{\varepsilon}
\newcommand{\R}{\mathbb{R}}

\newcommand{\vect}[1]{\boldsymbol{#1}}

\begin{document}
\title{Mechanics of axisymmetric sheets of interlocking and slidable rods}
\author{D. Riccobelli$^1$}
\email{davide.riccobelli@sissa.it}
\author{G. Noselli$^1$}
\email{giovanni.noselli@sissa.it}
\author{M. Arroyo$^{2,\,3,\,4}$}
\email{marino.arroyo@upc.edu}
\author{A. DeSimone$^{1,\,5}$}
\email{desimone@sissa.it}
\address[1]{SISSA -- International School for Advanced Studies, 34136 Trieste, Italy.}
\address[2]{LaC\`{a}N, Universitat Polit\`{e}cnica de Catalunya-BarcelonaTech, 08034 Barcelona, Spain.}
\address[3]{Institute for Bioengineering of Catalonia (IBEC), The Barcelona Institute of Science and Technology (BIST), 08028 Barcelona, Spain.}
\address[4]{Centre Internacional de M\`{e}todes Num\`{e}rics en Enginyeria (CIMNE), 08034 Barcelona, Spain.}
\address[5]{The BioRobotics Institute and Depertment of Excellence in Robotics and A.I., Sant’Anna School of Advanced Studies, 56127 Pisa, Italy.}

\begin{abstract}
In this work, we study the mechanics of metamaterial sheets inspired by the pellicle of Euglenids. They are composed of interlocking elastic rods which can freely slide along their edges. We characterize the kinematics and the mechanics of these structures using the special Cosserat theory of rods and by assuming axisymmetric deformations of the tubular assembly. Through an asymptotic expansion, we investigate both structures that comprise a discrete number of rods and the limit case of a sheet composed by infinitely many rods. We apply our theoretical framework to investigate the stability of these structures in the presence of an axial load. Through a linear analysis, we compute the critical buckling force for both the discrete and the continuous case. For the latter, we also perform a numerical post-buckling analysis, studying the non-linear evolution of the bifurcation through finite elements simulations.
\end{abstract}

\maketitle
\section{Introduction}

Mechanical metamaterials can exhibit unusual properties not present in the base material through a suitable design of their micro- or meso-architecture \citep{Bertoldi_2017}. Here, we examine the mechanics of a metamaterial consisting of sheets of interlocking and slidable rods forming a tubular surface.
Structures made of assemblies of rods are common and the geometrical constraints acting on the individual beams can significantly affect the mechanics of the assembly. The simplest example of geometrically interacting beams is provided by the birod, an assembly composed of two elastic beams glued together \citep{Moakher_2005}. This kinematic constraint can significantly affect the collective behavior of the two rods, leading to a complex mechanical behavior \citep{Lessinnes_2016}.
Other examples of interacting rod assemblies are provided by plies \citep{ neukirch2002geometry, Thompson_2002, Starostin_2014, Zhao_2014} and braided structures, such as McKibben pneumatic artificial muscles \citep{Tondu_2012} and biomimetic structures inspired by the motility of micro-organisms \citep{Cicconofri_2020}.

The material architecture we consider here is motivated by the cell wall (pellicle) of \emph{Euglena} cells \citep{handbook_protists,buetow1968biology}. The pellicle is a surface composed of adjacent flexible strips made of proteinaceous material. Pellicle strips are crosslinked by molecular motors, which can actively impose sliding between adjacent strips. By tuning the sliding in space and time, \emph{Euglena} cells can harmoniously change their shape to crawl in confined spaces \citep{Noselli2019a}. Similar to \cite{frenzel2017three}, we obtain a structure that twists under compressive loads, but thanks to a different geometry of the material micro-architecture. 

Most previous theoretical models of the pellicle are kinematical and establish a relation between the change in local metric achieved by sliding strips and shape  \citep{Arroyo2012,Arroyo_2014}, in the spirit of the theory of non-Euclidean plates and shells \citep{Sharon}. Exploiting the interplay between stretching and curvature encoded in the \emph{Theorema Egregium} of  Gauss is receiving increasing attention in the literature: we call ``Gaussian morphing''  this strategy of shape control of thin two-dimensional structures \citep{Cicconofri_2020}, which was pioneered in \citep{Sharon}. 
Envisaging an artificial realization of a morphable surface made of interlocking and slidable rods, \cite{Noselli_2019} developed a discrete mechanical model connecting the mechanics of the individual rods with that of the rod assembly, but that was restricted to shape transformations between cylinders.

In this paper, using mechanical modeling, asymptotic analysis and numerical calculations, we examine the nonlinear mechanics of axisymmetric sheets of interlocking and slidable rods focusing on the buckling and post-buckling behavior. 
We show that the kinematic constraints between adjacent rods fundamentally modify the load-bearing capacity of individual rods. The condition that two adjacent rods match along their common edge leads to a collective mechanical behavior which produces a coupling between shortening and twisting, and provides a means to tune (increase or decrease) the loading capacity of the structure depending on the bending and torsional stiffnesses of the rods, or on their spontaneous curvature.

\section{A sheet of interlocking and slidable  elastic rods}
\label{sec:model}
In this section, we develop a mathematical model of the sheet metamaterial under the condition of axisymmetry. The model is based on Kirchhoff rod theory \citep{antman_book, audoly2010elasticity} for each of the $n$ interlocking elastic rods, and supplemented by the kinematic constraints enforcing the interlocking and sliding connection between the rods. 

\subsection{Kinematics and compatibility constraint}

We consider $n$ identical Cosserat rods \citep{antman_book} having straight reference configuration and arranged on a cylinder of radius $R_0$ such that
\begin{equation}
\label{eq:initial_rods}
\left\{
\begin{aligned}
&\vect{r}_0^k(s)=R_0\cos\left(\frac{2k\pi}{n}\right)\vect{E}_X+R_0\sin\left(\frac{2k\pi}{n}\right)\vect{E}_Y+s\vect{E}_Z,\\[0.75mm]
&\vect{d}_{01}^k(s)=-\cos\left(\frac{2k\pi}{n}\right)\vect{E}_X-\sin\left(\frac{2k\pi}{n}\right)\vect{E}_Y,\\[0.75mm]
&\vect{d}_{02}^k(s)=\sin\left(\frac{2k\pi}{n}\right)\vect{E}_X-\cos\left(\frac{2k\pi}{n}\right)\vect{E}_Y,\\[0.75mm]
&\vect{d}_{03}^k(s)=\vect{E}_Z,
\end{aligned}
\right.\;
\end{equation}
with $s\in[0,L]$ and $k=0,\dots,n-1$. In Eq.~\eqref{eq:initial_rods}, 
$\vect{r}_0^k$ is the midline of the $k$-th rod in the reference configuration, $\vect{d}_{0i}^k$ is the $i$-th director of the $k$-th rod, $(\vect{E}_X,\,\vect{E}_Y,\,\vect{E}_Z)$ are the basis vectors of the reference space, and $L$ is the length of the rods. It is clear that $s$ is an arclength parameter for these curves, which we use as a Lagrangian coordinate. For concreteness, we assume that each rod has a rectangular cross-section and we denote by $t$ and $h$ its thickness and width in the $\vect{d}_{01}^k$ and  $\vect{d}_{02}^k$ directions, respectively. The director $\vect{d}_{03}^k$ is along the midline of the rods.

\begin{figure}[t!]
\centering
\includegraphics[height=0.6\textwidth]{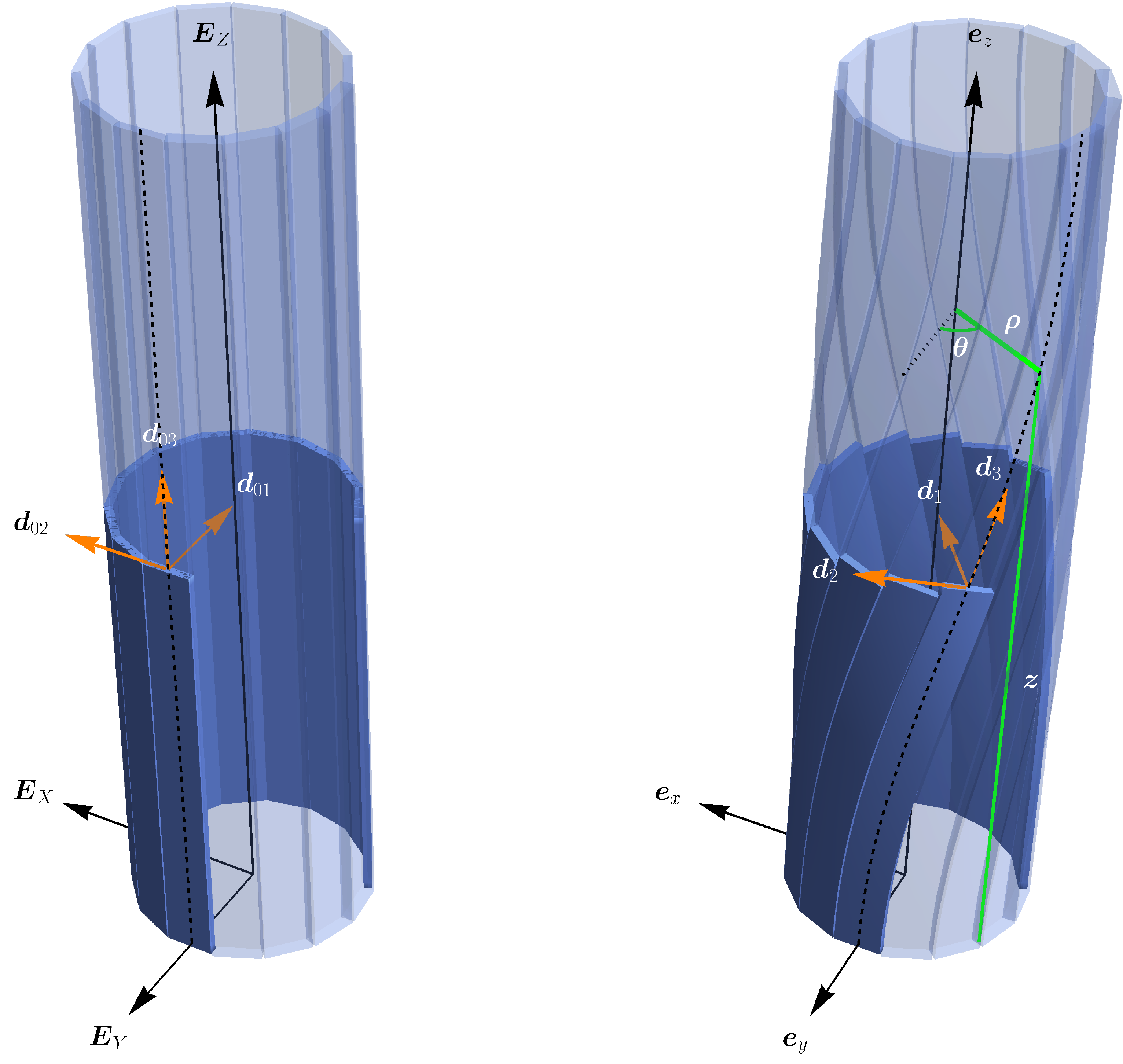}
\caption{Reference (left) and actual (right) representation of the rods assembly. Dashed lines denote the midlines of the rods.}
\label{fig:config}
\end{figure}

For $k=0$, the midline of the deformed rod can be described through a general parametrization of the curve in cylindrical coordinates in terms of the Lagrangian coordinate $s$, that is
\begin{equation}
\label{eq:rn}
\vect{r}_0(s)=\rho(s)\cos\theta(s)\vect{e}_x+\rho(s)\sin \theta(s) \vect{e}_y+z(s)\vect{e}_z,
\end{equation}
where $(\vect{e}_x,\,\vect{e}_y,\,\vect{e}_z)$ is the vector basis in the Eulerian reference,  $\rho$ is the radial distance from the symmetry axis $\vect{e}_z$, $\theta$ describes the change of the azimuthal angle, and $z$ is the actual height at $s$, see Fig.~\ref{fig:config}. By the assumption of axisymmetry, we can then parametrize the curve representing the deformed midline of the $k$\textsuperscript{th} rod as
\begin{equation}
\label{eq:r}
\vect{r}_k(s)=\rho(s)\cos\left(\frac{2k\pi}{n}+\theta(s)\right)\vect{e}_x+\rho(s)\sin\left(\frac{2k\pi}{n}+\theta(s)\right)\vect{e}_y+z(s)\vect{e}_z,
\end{equation}
and define the surface of revolution $\mathcal{S}$ generated by rotating $\vect{r}_0(s)$ about $\vect{e}_z$.
Assuming that rods are unshearable and inextensible, we obtain the constraints
\begin{equation}
\label{eq:d3}
\vect{r}_k'(s)=\vect{d}^k_3(s) ,
\end{equation}
in which a prime denotes differentiation with respect to the arclength parameter, and
\begin{equation}
\label{eq:inex}
|\vect{r}_k'(s)|=1,
\end{equation}
so that $s$ is the arclength of the midline both in the reference and in the actual configuration.

The director $\vect{d}^k_3(s)$ is tangent to the $k$\textsuperscript{th} rod, and thus, by the definition of $\mathcal{S}$, is also tangent to the surface at $\vect{r}_k(s)$. The other two directors in the deformed configuration, $\vect{d}_1^k$ and $\vect{d}_2^k$, are perpendicular to each other and belong to the plane orthogonal to $\vect{d}_3^k$, namely
\[
\vect{d}_1^k(s),\,\vect{d}_2^k(s)\in\mathcal{P}_s^k\qquad\text{where}\qquad\mathcal{P}_s^k =\left\{\vect{v}\in\R^3\;|\;\vect{d}_3^k(s)\cdot\vect{v}=0\right\}. 
\]

We construct next an orthonormal basis for the subspace $\mathcal{P}_s^k$ alternative to that formed by the first two directors. The first basis vector, $\vect{n}_k(s)$, is the inward normal to $\mathcal{S}$ along the $k$\textsuperscript{th} rod. We can express $\vect{n}_k$ as
\begin{equation}
\label{eq:v}
\vect{n}_k(s)=\frac{\vect{d}^k_3\times\vect{e}_\phi}{\left|\vect{d}^k_3\times\vect{e}_\phi\right|} ,
\end{equation}
where $\vect{e}_\phi$ is the unit vector representing the azimuthal direction in the cylindrical coordinate system.
Then, the second orthonormal vector $\vect{b}_k$ can be defined as
\begin{equation}
\label{eq:w}
\vect{b}_k\coloneqq\vect{d}^k_3\times\vect{n}_k.
\end{equation}

Since $(\vect{n}_k(s),\,\vect{b}_k(s), \vect{d}^k_3(s))$  and $(\vect{d}^k_1(s),\,\vect{d}^k_2(s), \vect{d}^k_3(s))$ are two right-handed orthonormal bases, there exists a rotation of axis $\vect{d}^k_3(s)$ and angle $\alpha(s)$
transforming one into the other. In other words, there exists a function $\alpha:(0,\,L)\mapsto\R$ such that
\begin{equation}
\label{eq:d1d2implicit}
\left\{
\begin{aligned}
&\vect{d}_1^k=\cos(\alpha)\vect{n}_k+\sin(\alpha)\vect{b}_k ,\\
&\vect{d}_2^k=-\sin(\alpha)\vect{n}_k+\cos(\alpha)\vect{b}_k .
\end{aligned}
\right.
\end{equation}

Substituting the expression of $\vect{r}_k$ given by Eq.~\eqref{eq:r} into Eq.~\eqref{eq:d3}, we can express $\vect{d}_3^k$ as a function of $\rho,\,\theta,\,z$ and their derivatives. By combining Eqs.~\eqref{eq:v}-\eqref{eq:d1d2implicit}, we can further express $\vect{d}_1^k$ and $\vect{d}_2^k$ in terms of $\rho,\,\theta,\,z$, and $\alpha$ and their derivatives, obtaining
\begin{gather}
\label{eq:d1_long}
\begin{aligned}
\vect{d}_1^k=
\frac{\sin (\alpha ) \left(\rho \rho' \theta ' \cos \left(\frac{2 \pi  k}{n}+\theta \right)+\left(\rho'^2+z'^2\right) \sin \left(\frac{2 \pi  k}{n}+\theta \right)\right)-\cos (\alpha ) z' \cos \left(\frac{2 \pi  k}{n}+\theta \right)}{\sqrt{\rho'^2+z'^2}}&\vect{e}_x+\\
+\frac{\sin (\alpha ) \left(\rho \rho' \theta ' \sin \left(\frac{2 \pi  k}{n}+\theta \right)-\left(\rho'^2+z'^2\right) \cos \left(\frac{2 \pi  k}{n}+\theta \right)\right)-\cos (\alpha ) z' \sin \left(\frac{2 \pi  k}{n}+\theta \right)}{\sqrt{\rho'^2+z'^2}}&\vect{e}_y+\\
+\frac{\rho' \cos (\alpha )+\rho \sin (\alpha ) \theta ' z'}{\sqrt{\rho'^2+z'^2}}&\vect{e}_z ,
\end{aligned}\\[1mm]
\label{eq:d2_long}
\begin{aligned}
\vect{d}_2^k=
\frac{\cos (\alpha ) \left(\rho \rho' \theta ' \cos \left(\frac{2 \pi  k}{n}+\theta \right)+\left(\rho'^2+z'^2\right) \sin \left(\frac{2 \pi  k}{n}+\theta \right)\right)+\sin (\alpha ) z' \cos \left(\frac{2 \pi  k}{n}+\theta \right)}{\sqrt{\rho'^2+z'^2}}&\vect{e}_x+\\
+\frac{\cos (\alpha ) \left(\rho \rho' \theta ' \sin \left(\frac{2 \pi  k}{n}+\theta \right)-\left(\rho'^2+z'^2\right) \cos \left(\frac{2 \pi  k}{n}+\theta \right)\right)+\sin (\alpha ) z' \sin \left(\frac{2 \pi  k}{n}+\theta \right)}{\sqrt{\rho'^2+z'^2}}&\vect{e}_y+\\
+\frac{\rho \cos (\alpha ) \theta ' z'-\rho' \sin (\alpha )}{\sqrt{\rho'^2+z'^2}}&\vect{e}_z ,
\end{aligned}\\[1mm]
\label{eq:d3_long}
\begin{aligned}
\vect{d}_3^k=\left(\rho' \cos \left(\frac{2 \pi  k}{n}+\theta \right)-\rho \theta ' \sin \left(\frac{2 \pi  k}{n}+\theta \right)\right)&\vect{e}_x+\\
+\left(\rho' \sin \left(\frac{2 \pi  k}{n}+\theta \right)+\rho \theta ' \cos
   \left(\frac{2 \pi  k}{n}+\theta \right)\right)&\vect{e}_y+ z'\vect{e}_z.
\end{aligned}
\end{gather}
Notice that, in view of Eq.~\eqref{eq:d3_long}, the inextensibility constraint given by Eq.~\eqref{eq:inex} becomes
\begin{equation}
\label{eq:inex_rho}
\rho'^2+\rho^2 \theta '^2+z'^2=1.
\end{equation}

\begin{figure}[t!]
\centering
\includegraphics[width=0.7\textwidth]{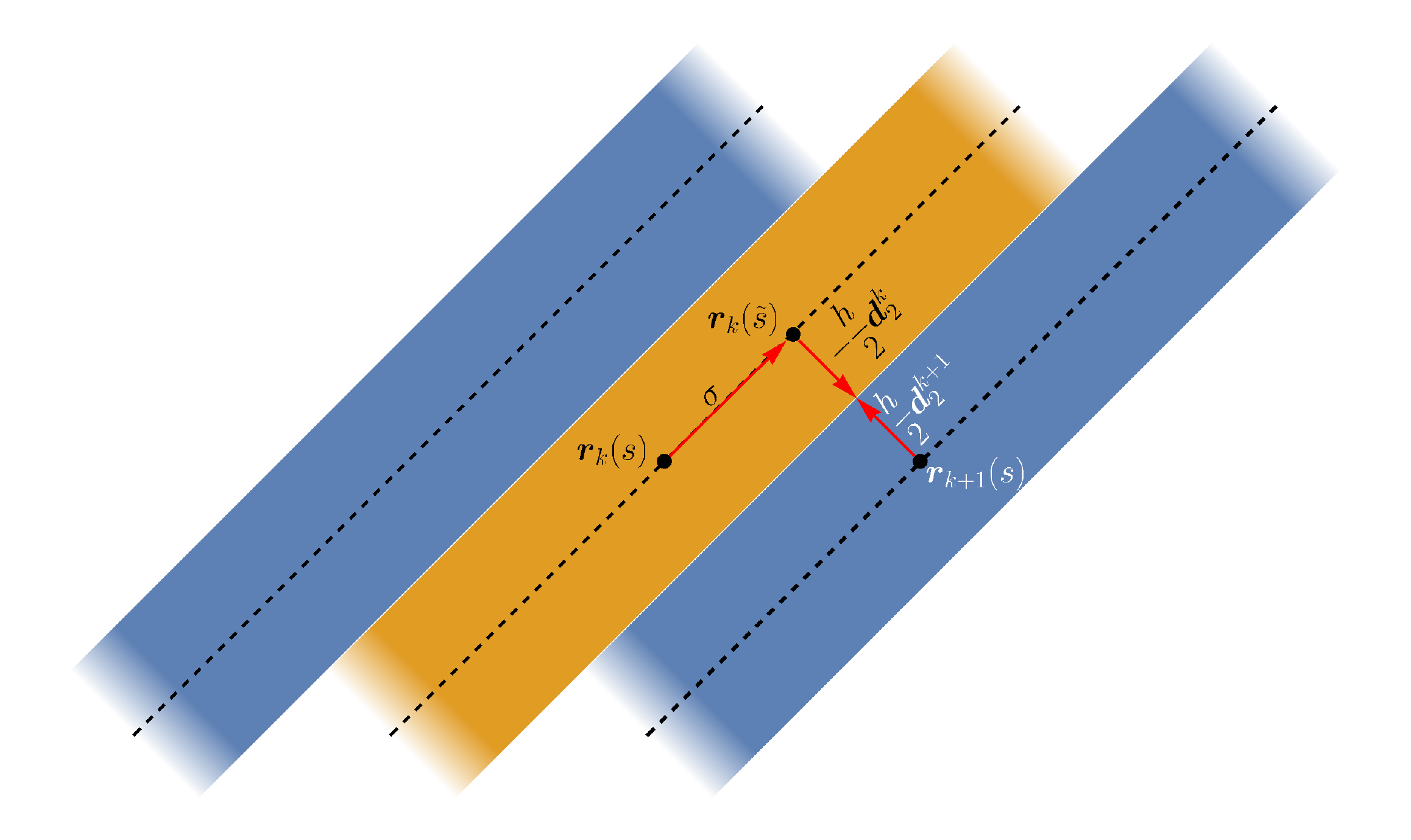}
\caption{A graphical representation of the compatibility constraint enforced by Eq.~\eqref{eq:compatibility}.}
\label{fig:compatibility}
\end{figure}

Up to now, we have required that the deformation of the rod assembly is axisymmetric but we have not enforced in our model the condition that the rods are interlocking and slidable. Following \cite{Noselli_2019}, this compatibility condition requires that the curves representing the edges of two adjacent rods coincide. Such a constraint is imposed by requiring that, for each point $P$ belonging to the common edge of the rods, there exist $s$ and $\tilde{s}$ belonging to $[0,L]$ such that:
\begin{equation}
\label{eq:compatibility}
\vect{r}_k(\tilde{s})-\frac{h}{2}\vect{d}_2^k(\tilde{s})=\vect{r}_{k+1}(s)+\frac{h}{2}\vect{d}_2^{k+1}(s),\qquad s\in[0,\,L],
\end{equation}
see Fig.~\ref{fig:compatibility} for a graphical representation of this constraint.
Compared with the kinematic constraint enforced on $n$ rods forming a ply \citep{neukirch2002geometry}, Eq.~\eqref{eq:compatibility} is rather different. In fact, in plies the cross-section is circular and the midlines lie at a fixed distance from adjacent rods. The compatibility equation \eqref{eq:compatibility} requires instead that neighboring beams must share a specific edge.
In this way, for each point $\vect{r}_{k+1}(s)$, having arclength $s$ on the $(k+1)$\textsuperscript{th} midline, we can identify a corresponding point $\vect{r}_k(\tilde{s})$ on the $k$\textsuperscript{th} midline, having arclength $\tilde{s}$. Through this correspondence, we can define a function $\tilde{s} = \tilde{s}(s)$ mapping $[0,\,L]$ onto itself. In the reference configuration $\tilde{s}(s)=s$, and since points along $\vect{r}_k$ cannot exchange ordering during  deformation, we have that $s_1<s_2$ implies that $\tilde{s}(s_1)<\tilde{s}(s_2)$. This shows that $\tilde{s}$ is strictly monotonic, and thus a bijection of $[0,\,L]$ onto $[0,\,L]$. Hence, we have that
\begin{equation}\label{aaa}
\tilde{s}(0) = 0,\qquad\tilde{s}(L) = L.
\end{equation}
We note that the compatibility condition enforced here is rather stringent, in that it holds over the entire common edge between adjacent rods. 
If we let the ends of the beams free to slide, the contact region over which the compatibility constraint holds is only a part of the interval $(0,\,L)$ and it depends on the configuration assumed by the assembly. Tracking the contact region then becomes a free boundary problem, which is beyond the scope of the present work. See \citep{Bigoni_2015, Cicconofri_2015} for related problems.

We finally introduce the function $\sigma:[0,\,L]\rightarrow\R$, defined as
\begin{equation}
\label{eq:sigma_def}
\sigma(s) = \tilde{s}(s) - s.
\end{equation}
Since in the reference configuration  $\tilde{s}(s)=s$, the function $\sigma(s)$ quantifies the (a-priori unknown) relative sliding between two adjacent rods. In view of \eqref{aaa}, we have
\begin{equation}\label{eq:BC_easy}
\sigma(0) = 0,\qquad\sigma(L) = 0.
\end{equation}

By enforcing the compatibility constraint of Eq.~\eqref{eq:compatibility} in the reference configuration,  we obtain that $\alpha$ and $\sigma$ are equal to zero and that
\begin{equation}
\label{eq:h}
h = 2R_0\tan\left(\frac{\pi}{n}\right),
\end{equation}
which provides a relation between $R_0,\,h$ and $n$, according to which $h$ is the length of one of the $n$ sides of a regular polygon circumscribed to a circumference of radius $R_0$. If we restrict the kinematics to cylindrical configurations such that $\sigma$ is uniform, we recover the model proposed in \citep{Noselli_2019} as shown in \ref{app:cylinders}. We note, however, that in the present theory  $\sigma$ is a function of the arclength $s$ and can be non-uniform along the rod.

With Eqs.~\eqref{eq:r} and \eqref{eq:d1_long}-\eqref{eq:d3_long} in mind, the compatibility equations \eqref{eq:compatibility} provide three scalar differential equations linking the five scalar functions $\rho,\,\theta,\,z,\,\alpha$ and $\sigma$. Noting that these functions also need to satisfy the inextensibility constraint of Eq.~\eqref{eq:inex_rho}, we are left with four differential equations for five unknowns. Thus, these equations should allows us, in principle, to express the functions $\rho,\,\theta,\,z,\,\alpha$ (which fully determine the rods midlines and directors) in terms of the sliding function $\sigma$, and arrive at a reduced theory formulated in terms of the single kinematic variable $\sigma$. While this program is too difficult to implement in general, it is feasible in two important special cases: i)~a linearized theory for finitely many rods in the limit of small sliding $\sigma$ and ii)~the continuum limit for an assembly of infinitely many strips of infinitesimal width, where a continuous shear variable $\gamma$ replaces  the discrete sliding $\sigma$. These will be addressed in Section~\ref{sec:small_sigma} and in Section~\ref{sec:infty}, respectively.

\subsection{Elasticity: constitutive assumptions}

We introduce the strain vector $\vect{u}_k$, defined through the relation
\[
{\vect{d}^k_j}' = \vect{u}_k\times\vect{d}^k_j.
\]
More explicitly, the strain components $u_j:=\vect{u}_k\cdot\vect{d}^k_j$ (no summation over repeated indices) are given by \citep{antman_book}
\[
u_1 = {\vect{d}^k_2}'\cdot\vect{d}_3^k,\qquad u_2 = {\vect{d}_3^k}'\cdot\vect{d}_1^k,\qquad u_3 = {\vect{d}_1^k}'\cdot\vect{d}_2^k,
\]
and represent two bending curvatures and the rate of twisting along the rod axis. Due to axisymmetry, the strain components $u_j$ do not depend on $k$, thus the subscript $k$ is omitted from now on. We next obtain an expression of the vector $\vect{u}$ as a function of $\rho,\,\theta$, $z$ and $\alpha$. From the expression of the directors given in Eqs.~(\ref{eq:d1_long}-\ref{eq:d3_long}), we find
\begin{equation}
\label{eq:u}
u_1 = \frac{ f_1 \sin(\alpha)+f_2 \cos(\alpha) }{\sqrt{\rho'^2+z'^2}},\qquad
u_2 = \frac{f_1 \cos(\alpha)-f_2 \sin(\alpha)  }{\sqrt{\rho'^2+z'^2}},\qquad
u_3 = \frac{f_3+\alpha'f_4+\sin (2 \alpha ) f_5}{2 \left(\rho'^2+z'^2\right)},
\end{equation}
where the dependence on $\alpha$ has been emphasized and
\begin{equation*}
\left\{
\begin{aligned}
f_1&= z' \left(\rho \theta '^2-\rho''\right)+\rho' z'',\\
f_2&=\rho \theta '' \left(\rho'^2+z'^2\right)+\theta ' \left(2 \rho'^3+\rho' \left(\rho \left(\rho \theta '^2-\rho''\right)+2 z'^2\right)-\rho z' z''\right),\\
f_3&= 2 \rho \rho'' \theta ' z'-2 \rho \rho' \theta ' z''+2 \rho'^2 \theta ' z'+2 \theta ' z'^3,\\
f_4&= \rho^2 \rho'^2 \theta '^2+\cos (2 \alpha ) \left(\rho'^2+z'^2\right) \left(\rho'^2+\rho^2 \theta '^2+z'^2-1\right)+2 \rho'^2 z'^2+\rho'^4+\rho'^2+\rho^2 \theta '^2 z'^2+z'^4+z'^2,\\
f_5&= \left(\rho'^2+z'^2\right) \left(\rho' \left(\rho''+\rho \theta '^2\right)+\rho^2 \theta ' \theta ''+z' z''\right).
\end{aligned}
\right.
\end{equation*}

We assume that the rods are elastic and postulate the existence of a function $\hat{\vect{m}}$ such that the torque $\vect{m}$ applied on the cross-section of a rod at $s$ is given by \citep{antman_book}
\[
\vect{m}(s) = \hat{\vect{m}}(s,\,\vect{u}).
\]
In particular, denoting by $m_j$ the component of $\vect{m}$ along $\vect{d}_j$, we assume the following constitutive relations
\begin{equation}
\label{eq:const_torque}
m_1 = B_1 (u_1 - u_1^*),\qquad m_2 = B_2 (u_2 - u_2^*),\qquad m_3 = T (u_3 - u_3^*),
\end{equation}
where $B_1, B_2$, and $T$ are bending and torsion elastic constant and $u_1^*$, $u_2^*$, and $u_3^*$ are the natural curvatures and twist of the rods. These constitutive relations derive from the  potential elastic energy of each rod given by
\begin{equation}
\label{eq:elastic_energy}
\mathcal{E} = \int_0^L W\,ds = \frac{1}{2}\int_0^L \left[ B_1 (u_1 - u_1^*)^2+ B_2 (u_2 - u_2^*)^2 + T (u_3 - u_3^*)^2\right]\,ds,
\end{equation}
where $W$ is the strain energy density per unit length.

We consider for definiteness rods with a rectangular cross-section having dimensions $t$ and  $h$ along the $\vect{d}_1$ and $\vect{d}_2$ directions, respectively, and let $t=\lambda h$.  In our analysis, we will typically consider the case of $\lambda< 1$, such that it is a classical results that
\begin{equation}
\label{eq:rigidities}
\left\{
\begin{aligned}
&B_1 = \frac{Eth^3}{12} = \frac{E\lambda h^4}{12}, \\
&B_2 = \frac{Eht^3}{12} = \frac{E\lambda^3 h^4}{12}, \\
&T = \frac{G\chi ht^3}{3} = \frac{G\chi \lambda^3h^4}{3} , 
\end{aligned}
\right.
\end{equation}
where $E$ and $G$ are Young's and shear modulus, respectively, and $\chi$ is a numerical factor depending on the aspect ratio $\lambda$ tending to $1$ when $\lambda$ tends to zero.
A possible simplification arises if we consider rods that are thin along $\vect{d}_1$, {\it i.e.} if $\lambda\ll1$. In this case, from Eq.~\eqref{eq:rigidities} we obtain that
\[
B_2 \ll B_1, \quad T \ll B_1.
\]
Under this assumption, that we call \emph{thin rod approximation}, we neglect the contributions of $u_2$ and $u_3$  to the strain energy density per unit length, which hence reduces to
\begin{equation}
\label{eq:thin_rod}
W = \frac{1}{2}B_1\left(u_1-u_1^*\right)^2 ,
\end{equation}
where the curvature $u_1$ is given by Eq.~\eqref{eq:u}.

Solving the fully non-linear equilibrium problem for our rod assembly would require finding the functions $\rho,\,\theta, z$ and $\alpha$ that make the energy functional \eqref{eq:elastic_energy} stationary, subjected to the inextensibility constraint \eqref{eq:inex} and the compatibility condition \eqref{eq:compatibility}, which introduces the additional unknown function $\sigma$. This set of nonlinear equation is complex to solve even numerically due to the non-local effects resulting from the compatibility constraint. In the next section, we study this system of non-linear ODEs through an asymptotic expansion about the reference configuration.

\section{Small deviation from the reference configuration}
\label{sec:small_sigma}

In this section, we characterize the compatible configurations of the cylindrical assembly under the assumption of small relative sliding $\sigma$ between adjacent rods. Let
\[
\varepsilon = \dfrac{\sup_{s\in(0,\,L)}\sigma}{R_0},
\]
so that we can write $\sigma = \varepsilon R_0 \tilde{\sigma}$ with $\sup_{s\in(0,\,L)}\tilde{\sigma}=1$.
We assume the following power series expansion:
\begin{equation}
\label{eq:second_order_expansion}
\left\{
\begin{aligned}
&\rho(s) = R_0 + \varepsilon \rho_1(s)+\varepsilon^2\rho_2(s)+o(\epsilon^2),\\
&\theta(s) = \varepsilon \theta_1(s)+\varepsilon^2\theta_2(s)+o(\epsilon^2),\\
&z(s) = s + \varepsilon z_1(s)+\varepsilon^2 z_2(s)+o(\epsilon^2),\\
&\alpha(s) = \varepsilon \alpha_1(s)+\varepsilon^2 \alpha_2(s)+o(\epsilon^2),
\end{aligned}
\right.\qquad\text{as }\epsilon\rightarrow 0.
\end{equation}

We impose that the constraints of Eqs.~\eqref{eq:inex_rho}-\eqref{eq:compatibility} are satisfied at each order in $\varepsilon$.
First, we perform a series expansion of the inextensibility constraint \eqref{eq:inex_rho} in $\varepsilon$ up to the second order, obtaining
\begin{equation}
\label{eq:inex_eps}
2 \epsilon  z_1'(s)+ \epsilon ^2 \left(R_0^2 \theta_{1}'(s)^2+\rho_{1}'(s)^2+z_1'(s)^2+2 z_2'(s)\right)+o(\varepsilon^2)=0.
\end{equation}
Assuming $z(0)=0$ and setting equal to zero the first term of \eqref{eq:inex_eps}, we obtain $z_1(s) = 0$. The $z$ component of the compatibility equation~\eqref{eq:compatibility} at order $\epsilon$ reads
\[
R_0 \tilde{\sigma}(s)-2 R_0^2 \tan \left(\frac{\pi }{n}\right) \theta_{1}'(s)=0.
\] 
Thus, assuming $\theta(0)=0$, we obtain
\[
\theta_1(s) = \frac{1}{2R_0}\cot \left(\frac{\pi }{n} \right)\int_0^s \tilde{\sigma}(\tau)\,d\tau.
\]
We can now compute the first order of the series expansion of the radial and azimuthal component of the compatibility equation. Setting them equal to zero we obtain
\[
\left\{
\begin{aligned}
\rho_{1}&=-R_0\tan \left(\frac{\pi }{n}\right) \alpha_{1} ,\\
\rho_{1}&= R_0\cot \left(\frac{\pi }{n}\right) \alpha_{1} ,
\end{aligned}
\right.
\]
and thus $\alpha_1=\rho_1 = 0$.

The computation can be iterated at the second order following the same procedure. First, we compute $z_2$ from the inextensibility equation \eqref{eq:inex_eps}
\[
z_2(s) = -\frac{1}{8} \cot ^2\left(\frac{\pi }{n}\right)\int_0^s \tilde{\sigma}(\alpha )^2\,ds.
\]
Second, we project the compatibility equation along the $z$ direction
\[
-\frac{1}{2} R_0^2 \left(4 \tan \left(\frac{\pi }{n}\right) \theta_{2}'+\tilde{\sigma} \tilde{\sigma}'\right)\epsilon^2+o(\varepsilon^2)=0,
\]
and thus
\[
\theta_2(s) = \frac{1}{8} \cot \left(\frac{\pi }{n}\right) \left(\tilde{\sigma}(0)^2-\tilde{\sigma}(s)^2\right).
\]
Finally, we obtain $\rho_2$ and $\alpha_2$ projecting the compatibility equation along the radial and the azimuthal direction. The expressions at the second order in $\varepsilon$ read
\[
\left\{
\begin{aligned}
-&2 R_0 \sin ^2\left(\frac{\pi }{n}\right) \alpha_{2}-\frac{1}{8} R_0 \left(\cos \left(\frac{2 \pi }{n}\right)-3\right) \cot \left(\frac{\pi }{n}\right) \tilde{\sigma}^2-\sin \left(\frac{2 \pi }{n}\right) \rho_{2}=0 ,\\
-&R_0 \sin \left(\frac{2 \pi }{n}\right) \alpha_{2}+\frac{1}{8} R_0 \left(\cos \left(\frac{2 \pi }{n}\right)-3\right) \tilde{\sigma}^2+2 \sin ^2\left(\frac{\pi }{n}\right) \rho_{2}=0 ,
\end{aligned}
\right.
\]
from which we obtain
\begin{equation}
\label{eq:secondorderralpha}
\left\{
\begin{aligned}
&\rho_{2}= \frac{1}{8} R_0 \left(\left(\csc ^2\left(\frac{\pi }{n}\right)+1\right) \tilde{\sigma}^2-8 \tan \left(\frac{\pi }{n}\right) \alpha_{2}\right),\\
&\rho_{2}=R_0 \cot \left(\frac{\pi }{n}\right) \alpha_{2}+\frac{1}{8} R_0 \left(\csc ^2\left(\frac{\pi }{n}\right)+1\right) \tilde{\sigma}^2 ,
\end{aligned}
\right.
\end{equation}
a system whose solution is given by
\[
\left\{
\begin{aligned}
&\rho_2 = \frac{R_0}{8} \left(\csc ^2\left(\frac{\pi }{n}\right)+1\right) \tilde{\sigma}^2,\\
&\alpha_2 = 0.
\end{aligned}
\right.
\]

Summarizing these results, the asymptotic expansion shows that
\begin{equation}
\label{eq:svilupporho}
\left\{
\begin{aligned}
\rho(s) &= R_0 +\frac{\epsilon ^2}{8} R_0  \left(\csc ^2\left(\frac{\pi }{n}\right)+1\right) \tilde\sigma(s)^2+o(\varepsilon^2),\\
\theta(s) &=  \frac{\varepsilon}{2 R_0} \cot \left(\frac{\pi }{n}\right)\int_0^s\tilde\sigma(\tau)\,d\tau+\frac{\varepsilon^2}{8} \cot \left(\frac{\pi }{n}\right)(\tilde\sigma (0)^2-\tilde\sigma (s)^2)+o(\varepsilon^2),\\
z(s) &= s - \frac{\varepsilon^2}{8} \cot ^2\left(\frac{\pi }{n}\right)\int_0^s \tilde\sigma(\tau)^2\,d\tau+o(\varepsilon^2),\\
\alpha(s) &= o(\varepsilon^2).
\end{aligned}
\right.\qquad\text{as }\epsilon\rightarrow 0.
\end{equation}
The fact that $\alpha_1$ and $\alpha_2$ are both zero raises the question of whether $\alpha$ is identically zero. In \ref{app:third_order}, we compute the third-order expansion and show that this is not the case.

Finally, we express the curvatures and the torsion of the rod in terms of $\sigma$ by substituting  Eq.~\eqref{eq:svilupporho} into Eq.~\eqref{eq:u}, obtaining
\begin{equation}
\label{eq:u_small_sigma}
\left\{
\begin{aligned}
u_1 &= \frac{1}{2} \epsilon  \cot \left(\frac{\pi }{n}\right) \tilde{\sigma}'(s)-\frac{\epsilon^2R_0}{4}\cot\left(\frac{\pi}{n}\right) \left( \tilde{\sigma}(s) \tilde{\sigma}''(s)+ \tilde{\sigma}'(s)^2\right)+o(\epsilon^2),\\
u_2 &=\epsilon ^2 \left(\frac{\cot ^2\left(\frac{\pi }{n}\right) \tilde{\sigma}^2}{4 R_0}-\frac{1}{4} R_0\left(\csc ^2\left(\frac{\pi }{n}\right)+1\right) \tilde{\sigma} \tilde{\sigma}''-\frac{1}{4} R_0\left(\csc ^2\left(\frac{\pi }{n}\right)+1\right) \tilde{\sigma}'^2\right)+o(\epsilon^2),\\
u_3 &=\frac{\epsilon  \cot \left(\frac{\pi }{n}\right) \tilde{\sigma}}{2 R_0}-\frac{1}{4} \epsilon ^2 \left(\cot \left(\frac{\pi }{n}\right) \tilde{\sigma} \tilde{\sigma}'\right)+o\left(\epsilon ^2\right).
\end{aligned}
\right.
\end{equation}
Using these formulae and ignoring higher-order terms, we can express the elastic energy~\eqref{eq:elastic_energy} as a functional of the sliding function $\sigma$ only
\begin{equation}
\label{eq:energy_sigma}
\mathcal{E}[\sigma] = \frac{1}{2}\int_0^L \left[ B_1 (u_1(\sigma) - u_1^*)^2+ B_2 (u_2(\sigma) - u_2^*)^2 + T (u_3(\sigma) - u_3^*)^2\right]\,ds.
\end{equation}
In the following section, we perform an analogous reduction of the energy functional \eqref{eq:elastic_energy} in terms of a single function by considering the limit case of an infinite number of rods.

\section{Limit case of an infinite number of rods}\label{sec:infty}

Another interesting situation amenable to explicit calculations is the limit of an infinite number of rods, $n\rightarrow\infty$. Remarkably, \cite{Noselli_2019} showed that this limit was a good approximation of the discrete case for $n\geq 5$.
Equation \eqref{eq:h} shows that $h=O(n^{-1})$. As the number of rods blows up, it is reasonable to expect that the sliding displacement between infinitesimally narrow rods will become small. It is  thus natural to define 
\[
\gamma(s)=\frac{\sigma(s)}{h},
\]
which can be interpreted as an in-plane shear strain of the rod assembly along the rod direction.

We assume the following asymptotic expansions for large $n$
\begin{equation}
\label{eq:asympt_exp_n_inf}
\left\{
\begin{aligned}
&\rho(s) = \rho_\infty(s) + o(1),\\
&\theta(s) = \theta_\infty(s) + o(1),\\
&z(s) = z_\infty(s) + o(1),\\
&\alpha(s) = \alpha_\infty(s) + o(1),
\end{aligned}
\right.\qquad\text{as } n\rightarrow+\infty.
\end{equation}
We now obtain an expression of $\rho_\infty$,  $\theta_\infty$,  $z_\infty$, and $\alpha_\infty$ as a function of $\gamma$ by enforcing the compatibility and the inextensibility constraints in Eqs.~\eqref{eq:compatibility} and \eqref{eq:inex}. First, we project the compatibility equation \eqref{eq:compatibility} along the $z$ direction. Noting that $\sigma=h \gamma$, we perform a Taylor series expansion about $n = \infty$ to obtain
\[
\frac{2 \pi  R_0 \left(\rho_\infty' \sin (\alpha_\infty )+z_\infty' \left(\gamma  \sqrt{\rho_\infty'^2+z_\infty'^2}-\rho_\infty \cos (\alpha_\infty ) \theta_\infty '\right)\right)}{n \sqrt{\rho_\infty'^2+z_\infty'^2}}+o\left(\frac{1}{n}\right)=0,
\]
and hence
\begin{equation}
\label{eq:gamma}
\gamma=\frac{\rho_\infty \cos (\alpha_\infty ) \theta_\infty ' z_\infty'-\rho_\infty' \sin (\alpha_\infty )}{z_\infty' \sqrt{\rho_\infty'^2+z_\infty'^2}}+o(1).
\end{equation}
Performing an expansion of the compatibility equations projected along the azimuthal direction, we obtain
\[
-\frac{2 \pi  R_0 \sin (\alpha_\infty ) \sqrt{\rho_\infty'^2+z_\infty'^2}}{n z_\infty'}+o\left(\frac{1}{n}\right)=0.
\]
Since $\rho_\infty'^2+z_\infty'^2\neq0$ (if otherwise $\gamma\to\infty$), we obtain 
\begin{equation}
\alpha_\infty = 0.
\end{equation}
From the inextensibility constraint \eqref{eq:inex_rho}, at order zero in $n^{-1}$ we obtain
\begin{equation}
\label{eq:z'ninf}
z_\infty' = \pm\sqrt{-\rho_\infty'^2-\rho_\infty^2 \theta_\infty '^2+1}.
\end{equation}
The plus and minus solutions are related by symmetry and in the following we choose the positive one. Finally, we project the compatibility equation \eqref{eq:compatibility} along the radial direction, obtaining
\begin{equation}
\label{eq:comp_3_n_inf}
\frac{2 \pi}{n}  \left(\frac{R_0}{\sqrt{1-\rho_\infty^2 \theta_\infty '^2}}-\rho_\infty\right)+o\left(\frac{1}{n}\right)=0.
\end{equation}
We solve Eqs.~\eqref{eq:gamma} and \eqref{eq:comp_3_n_inf} at the leading order in $n^{-1}$ with respect to $\rho_\infty$ and $\theta_\infty'$. Together with Eq.~\eqref{eq:z'ninf}, we obtain
\begin{equation}
\label{eq:rtz_n_infinity}
\left\{
\begin{aligned}
&\rho_\infty = R_0 \sqrt{\gamma ^2+1},\\
&\theta_\infty' = \frac{\gamma }{R_0 \left(\gamma ^2+1\right)},\\
&z_\infty' = \sqrt{\frac{1-R_0^2 \gamma ^2 \gamma '^2}{\gamma ^2+1}},\\
&\alpha_\infty=0.
\end{aligned}
\right.
\end{equation}

Interestingly,  Eq.~\eqref{eq:rtz_n_infinity} are identical to Eqs.~(9)-(11) in \citep{Arroyo_2014}, who studied the kinematics of a continuous cylindrical surface that could experience simple in-plane shear deformation along a Lagrangian direction, but ignoring the meso-scale architecture of the assembly. Thus, the present study shows that the theory developed by~\cite{Arroyo_2014} arises as the limit for $n\to\infty$ of the discrete theory for assemblies of $n$ rods sliding along the common edge that is studied here. Remarkably, our study shows that in the limit of large $n$, which as shown by \cite{Noselli_2019}  is an accurate approximation even for an assembly with a relatively small number of rods, $\alpha_\infty = 0$ and hence it is justified to assume that the director $\vect{d}_1^k$ remains normal to the surface $\mathcal{S}$ defined by the rod assembly. 

Analogously to the previous section, the rod strains $\vect{u}$ can also be explicitly expressed as a function of the local shear $\gamma$ thanks to Eqs.~\eqref{eq:u}-\eqref{eq:rtz_n_infinity} as
\begin{equation}
\label{eq:u_n_inf}
\left\{
\begin{aligned}
&u_1 = \gamma ' + o\left(\frac{1}{n}\right),\\
&u_2 = \frac{\gamma ^2 \left(1-R_0^2 \left(\gamma ^2+1\right) \gamma '^2\right)-R_0^2 \left(\gamma ^3+\gamma\right) \gamma ''-R_0^2 \gamma '^2}{R_0 \left(\gamma ^2+1\right)^{3/2} \sqrt{1-R_0^2 \gamma ^2 \gamma '^2}} + o\left(\frac{1}{n}\right),\\
&u_3 = \frac{\gamma  \left(R_0^2 \left(\left(\gamma ^3+\gamma \right) \gamma ''+\gamma '^2\right)+1\right)}{R_0 \left(\gamma ^2+1\right)^{3/2} \sqrt{1-R_0^2 \gamma ^2 \gamma '^2}} + o\left(\frac{1}{n}\right).
\end{aligned}
\right.
\end{equation}
Substituting Eq.~\eqref{eq:u_n_inf} into Eq.~\eqref{eq:elastic_energy} and neglecting the remainder, we can express the elastic energy of a rod in the assembly as a functional of the shear strain $\gamma$
\begin{equation}
\label{eq:energy_gamma}
\mathcal{E}[\gamma] = \frac{1}{2}\int_0^L \left[ B_1 (u_1(\gamma) - u_1^*)^2+ B_2 (u_2(\gamma) - u_2^*)^2 + T (u_3(\gamma) - u_3^*)^2\right]\,ds.
\end{equation}
Thus, both in the case of small $\sigma$ and of $n\rightarrow+\infty$, we have been able to write the elastic potential energy as a function of a single unknown function, $\sigma$ and $\gamma$ respectively, see Eqs.~\eqref{eq:energy_sigma} and \eqref{eq:energy_gamma}.

\section{Stability analysis of a cylindrical assembly subjected to an axial load}
\label{sec:buckling}

We apply our two theories, for small relative sliding and for large number of rods, developed in the previous sections to study the stability of a cylindrical rod assembly subjected to an axial force. For the sake of simplicity, we assume that the natural strains $u_i^*$ are constant.

\subsection{Linear stability analysis using the discrete model}

In this section, we use Eq.~\eqref{eq:u_small_sigma} to characterize the bending and the torsion of the rod. As discussed in Section~\ref{sec:small_sigma}, these equations allow us to write the potential energy of the rod as a function of $\sigma$ only, see Eq.~\eqref{eq:energy_sigma}.

\subsubsection{Case (a): Free rotation}
\label{sec:lin_stab_n_fin}
We assume that each rod composing the structure is loaded at $s=L$ with a force
\[
\vect{F} = F\vect{e}_z,
\]
so that its work is given by
\[
\mathcal{P}[\sigma] = \vect{F}\cdot(\vect{r}_k(L)-\vect{r}_0^k(L))=F (z(L)-L).
\]
The total energy of the mechanical system reads
\begin{equation}
\label{eq:total_energy_force}
\Psi[\sigma] = \mathcal{E}[\sigma]- \mathcal{P}[\sigma],
\end{equation}
and, using Eqs. \eqref{eq:svilupporho}-\eqref{eq:u_small_sigma}, its power series expansion about $\varepsilon=0$ is
\begin{equation}
\label{eq:total_energy_expansion}
\Psi = \Psi_0 + \varepsilon \Psi_1 + \frac{\varepsilon^2}{2}\Psi_2 + o(\varepsilon^2).
\end{equation}
The term of order zero is given by
\begin{equation}
\label{eq:Psi0}
\Psi_0 = \frac{1}{2} \left(B_1{u_1^*}^2+B_2{u_1^*}^2+T {u_3^*}^2\right)L,
\end{equation}
which is a constant independent of $\tilde{\sigma}$. As for the term of order $\epsilon$ in Eq.~\eqref{eq:total_energy_expansion}, this reads
\[
\begin{aligned}
\Psi_1[\tilde{\sigma}] &=-\cot \left(\frac{\pi }{n}\right) \int_0^L  
\frac{ T u_3^*\tilde{\sigma} + B_1 R_0 u_1^*\tilde{\sigma}' }{2 R_0}\,ds=
-\cot \left(\frac{\pi }{n}\right)\int_0^L \frac{T u_3^*\tilde{\sigma}}{2 R_0}\,ds,
\end{aligned}
\]
where the term multiplying $\tilde{\sigma}'$ under the integral sign disappears due to the boundary conditions $\sigma(0) = \sigma(L) = 0$. We observe that $\Psi_1$ corresponds to the first variation of Eq.~\eqref{eq:total_energy_force} about the reference configuration, {\it i.e.}, $\Psi_1[\tilde{\sigma}] = \delta\Psi(0)[R_0\tilde{\sigma}]$. This is zero for all the admissible $\tilde\sigma$ if and only if
\begin{equation}
\label{eq:no_torsion}
T u_3^* = 0,
\end{equation}
which is the equilibrium condition for the straight configuration of the cylindrical assembly. By accounting for the results of Eq. \eqref{eq:no_torsion}, the term $\Psi_2$ is a functional of $\tilde{\sigma}$ and its expression reads
\begin{equation}
\label{eq:Psi2}
\begin{aligned}
\Psi_2[\tilde{\sigma}]=\int_0^L\psi_2(\tilde{\sigma})\,ds & = \frac{1}{4R_0^2} \int_0^L\left[2 R_0^2 \tilde{\sigma} \left(R_0 \tilde{\sigma}'' \left(B_1 u_1^* \cot \left(\frac{\pi }{n}\right)+B_2 u_2^* \left(\csc ^2\left(\frac{\pi }{n}\right)+1\right)\right)
\right)+\right.\\
&+R_0^2\tilde{\sigma}'^2 \left(\csc ^2\left(\frac{\pi }{n}\right) (B_1+2 B_2 R_0 u_2^*)+2 B_1 R_0 u_1^* \cot \left(\frac{\pi }{n}\right)-B_1+2 B_2 R_0 u_2^*\right)+\\
&\left.+\cot ^2\left(\frac{\pi }{n}\right)\tilde{\sigma}^2 \left(T+F R_0^2-2 B_2 R_0 u_2^*\right)\right]\,ds,
\end{aligned}
\end{equation}
where $\psi_2$ is the integrand of the functional. 
To study the stability of the reference configuration, we exploit the Trefftz criterion and compute the first variation of $\Psi_2$ with respect to $\tilde{\sigma}$ \citep{Ba_ant_2010}.
\begin{figure}[t!]
\centering
\includegraphics[align=c, height=0.27\textwidth]{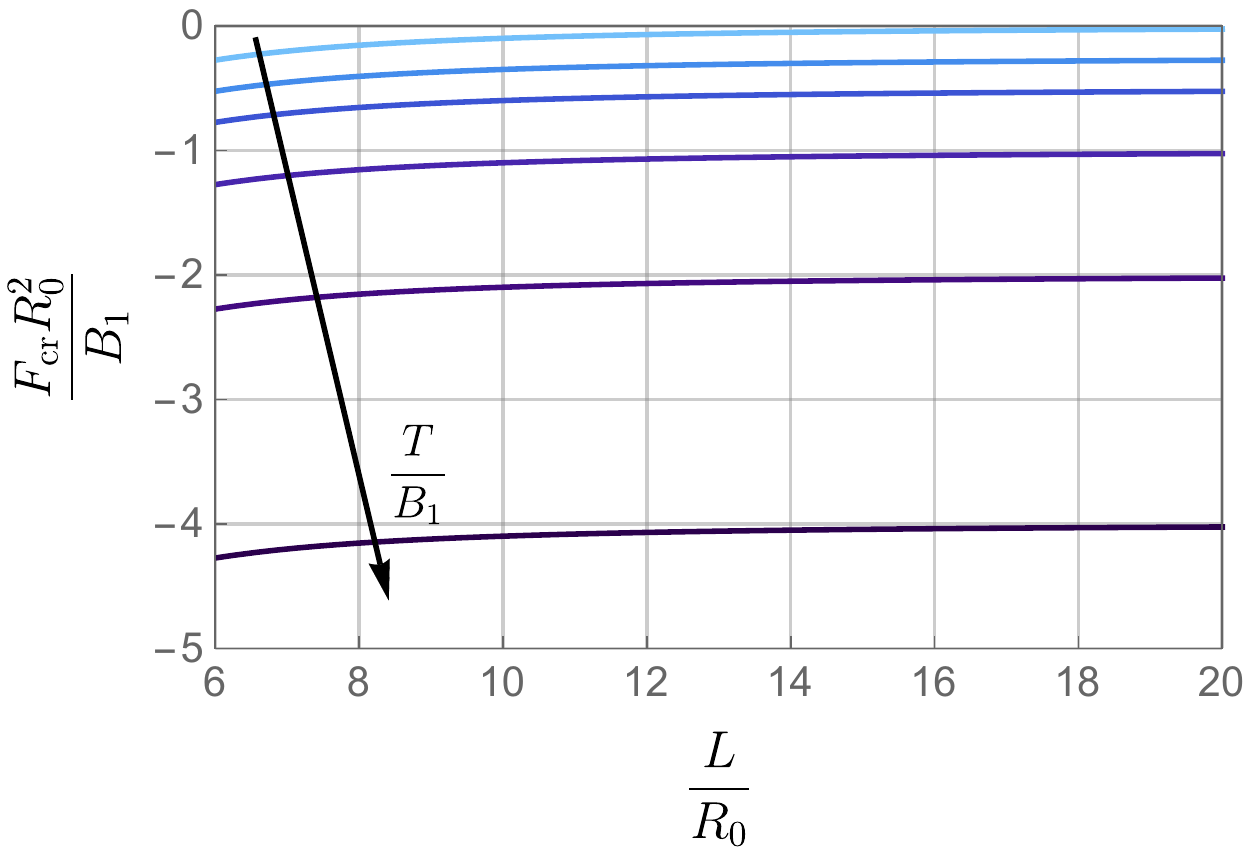}
\includegraphics[align=c, height=0.27\textwidth]{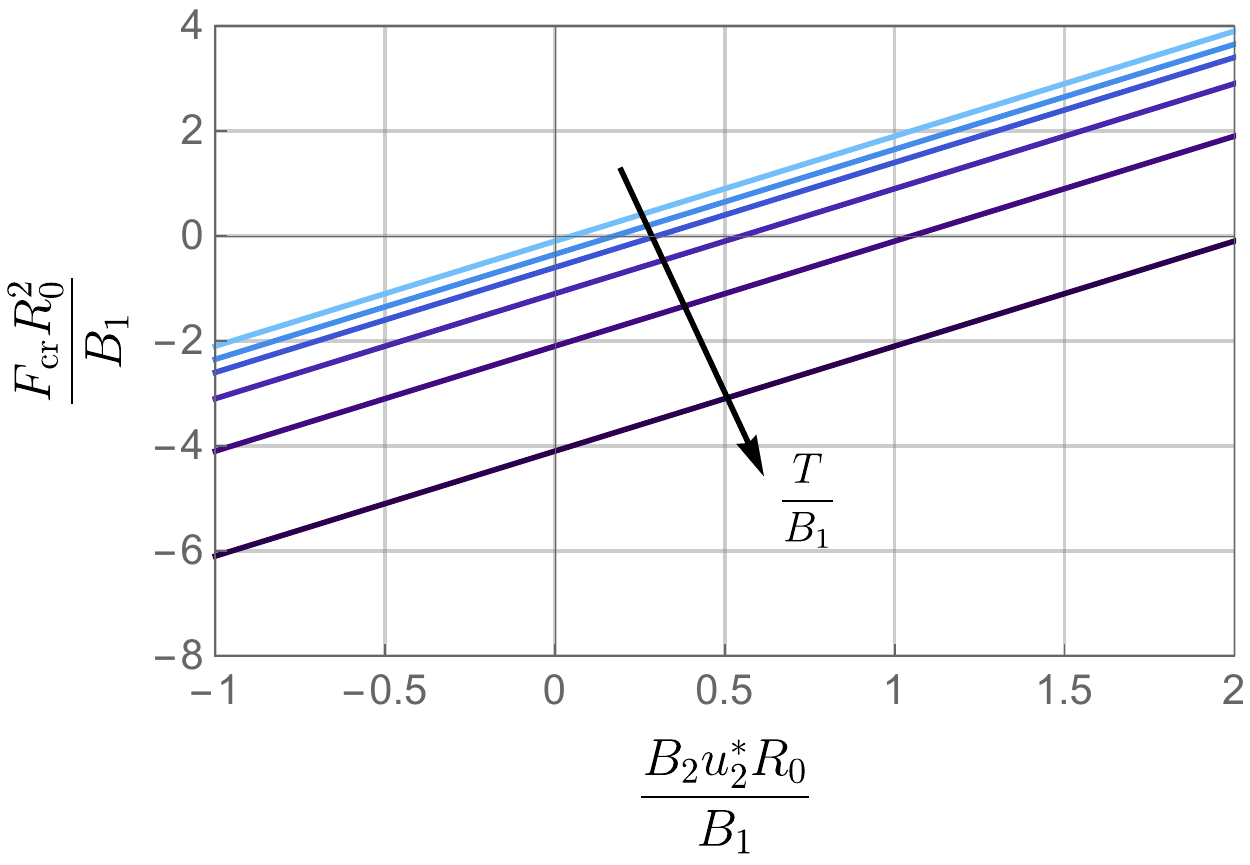}
\includegraphics[align=c, height=0.4\textwidth]{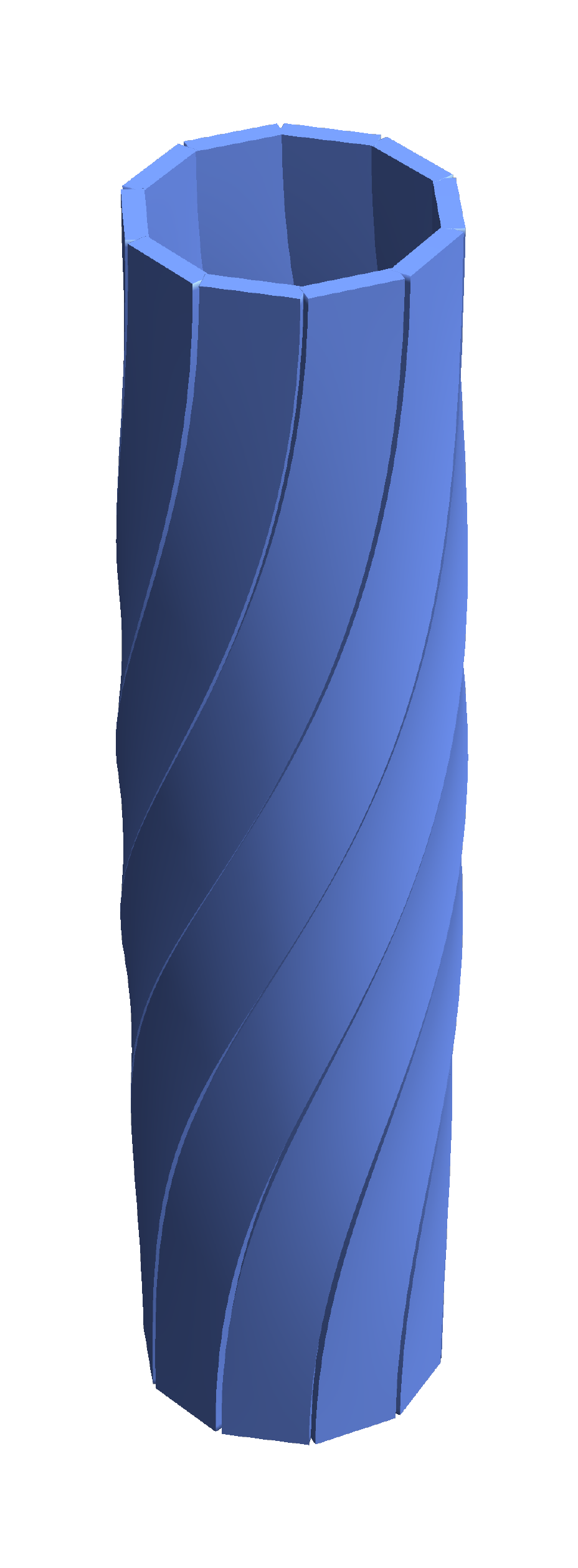}
\caption{Non-dimensional critical buckling load $F_\text{cr}R_0^2/B_1$ versus $L/R_0$ when $u_2^*=0$ (left) and $F_\text{cr}R_0^2/B_1$ versus $B_2 u_2^*R_0/B_1$ when $L/R_0=10$ (center) for $T/B_1 = 0,\,1/4,\,1/2,\,1,\,2,\,4$. The arrows denote the direction in which $T/B_1$ grows. (Right)
bifurcation mode for $n=10$ and $L/R_0 = 10$. The amplitude of $\sigma$ has been set to $0.25 R_0$ for the sake of graphical illustration.}
\label{fig:clamped}
\end{figure}
According to this criterion, a bifurcation occurs if there exists a function $\tilde{\sigma}$ such that
\begin{equation}
\label{eq:Trefftz}
\delta\Psi_2(\tilde{\sigma})[\delta\sigma]=\int_0^L\left[\frac{\partial\psi_2}{\partial\tilde{\sigma}}-\frac{d}{ds}\frac{\partial \psi_2}{\partial\tilde{\sigma}'}+\frac{d^2}{ds^2}\frac{\partial \psi_2}{\partial\tilde{\sigma}''}\right]\delta\sigma\,ds+\left[\frac{\partial \psi_2}{\partial\tilde{\sigma}''}\delta\sigma'\right]_0^L = 0,
\end{equation}
for any admissible $\delta\sigma$.
The natural boundary conditions arising from Eq. \eqref{eq:Trefftz} are given by
\begin{equation}
\label{eq:boundary_term}
\frac{\partial \psi_2}{\partial\tilde{\sigma}''}=\frac{1}{2} R_0 \tilde{\sigma}(s) \left(B_1 u_1^* \cot \left(\frac{\pi }{n}\right)+B_2 u_2^* \left(\csc ^2\left(\frac{\pi }{n}\right)+1\right)\right) = 0,\qquad\text{for } s=0,\,L,
\end{equation}
which are identically satisfied by the condition $\tilde{\sigma}(0)= \tilde{\sigma}(L)=0$ resulting from our edge compatibility condition. As regards the Euler--Lagrange equation, this reads
\begin{equation}
\label{eq:EL_disc}
B_1 R_0^2 \tilde{\sigma}''+\tilde{\sigma} \left(2 B_2 R_0 u_2^*-F R_0^2-T\right)=0 .
\end{equation}

We are now in the position to study the stability of the reference configuration. Let
\[
\tilde{\sigma}_m(s) = \sin\left(\frac{m\pi}{L}s\right),\qquad m\in\mathbb{N}\setminus\{0\},
\]
where $m$ is the wavenumber. The mode $\tilde{\sigma}_m$ is a solution of the Euler-Lagrange equation \eqref{eq:EL_disc} whenever
\[
F = F_m = -\frac{\pi ^2 B_1 m^2}{L^2}-\frac{T-2 B_2 R_0 u_2^*}{R_0^2}\qquad\text{and}\qquad u_3^* = 0.
\]
The critical wavenumber is $m = 1$ for a critical load of
\begin{equation}
\label{Fcr_a}
F_\text{cr} = F_1 = -\frac{\pi ^2 B_1}{L^2}-\frac{T-2 B_2 R_0 u_2^*}{R_0^2}.
\end{equation}

The configuration corresponding to this critical mode is reported in Fig.~\ref{fig:clamped} (right), which involves rotation of one of the ends of the assembly with respect to the other end. 
Interestingly, the fist term of the critical load coincides with that of a compressed and isolated rod that buckles by bending along the $\vect{d}_1^k$ direction with constrained rotation at both ends but free lateral translation, a situation analogous to the rods in the assembly. However, rods in the assembly must twist as they deform, which increases the critical load with the term proportional to $T$. This term is divided by $R_0$, which becomes the parameter determining slenderness for this contribution as opposed to $L$ for the first term. Finally, natural curvature in the normal direction to the surface (thus, perpendicular to the main buckling direction) can either increase or decrease the critical load depending on the sign, see Fig.~\ref{fig:clamped} (left and center). These results thus show how the constraints imposed by the rod assembly modify the buckling behavior of the rods, and how these constraints allow us to tune buckling by tuning independently the bending and twisting stiffnesses, natural curvatures and the dimensions of the assembly.

\subsubsection{Case (b): Constrained rotation}
\label{sec:lin_stab_const_n_fin}
We further constrain the system by not allowing the relative rotation of the cylinder ends during the deformation. Mathematically, this requirement is imposed by
\begin{equation}
\label{eq:no_rotation}
\theta(0) = \theta(L).
\end{equation}
We introduce a Lagrange multiplier $p$ to impose such a constraint, {\it i.e.}, we modify the total energy in Eq.~\eqref{eq:total_energy_force} by adding the term
\[
p (\theta(L)-\theta(0)),
\]
to obtain
\begin{equation}
\label{eq:energy_no_rotation_disc}
\Psi[\sigma,\,p] = \mathcal{E}[\sigma] - \mathcal{P}[\sigma] + p (\theta(L)-\theta(0)).
\end{equation}

Under the assumption of small relative sliding, we can use Eq.~\eqref{eq:svilupporho} to compute the expansion of $\theta(L)-\theta(0)$. Furthermore, we assume the following expansion for the Lagrange multiplier
\[
p = p_0 +\varepsilon p_1 + o(\varepsilon),
\]
where $p_0$ is the Lagrange multiplier in the reference configuration and $p_1$ is its correction as due to small sliding $\sigma$.

Following the same procedure as in Section \ref{sec:lin_stab_n_fin}, we compute the expansion of the Lagrangian \eqref{eq:energy_no_rotation_disc} up to the second order in $\varepsilon$, obtaining
\begin{equation}
\label{eq:expand_Psi_no_rot}
\Psi = \Psi_0 + \varepsilon \Psi_1 + \frac{\varepsilon^2}{2} \Psi_2 + o(\varepsilon^2),
\end{equation}
where the term of order zero is given by Eq. \eqref{eq:Psi0}, so that $\Psi_0$ is  a constant independent of $\tilde{\sigma},\,p_0$, and $p_1$. The term of order $\epsilon$ in Eq.~\eqref{eq:expand_Psi_no_rot} reads
\[
\begin{aligned}
\Psi_1[\tilde{\sigma}, \, p_0] &= \int_0^L \cot \left(\frac{\pi }{n}\right) 
\left(\frac{ p_{0} \tilde{\sigma} - T u_3^*\tilde{\sigma} - B_1 R_0 u_1^*\tilde{\sigma}' }{2 R_0}\right)ds=
\int_0^L \cot \left(\frac{\pi }{n}\right) \left(\frac{ p_{0}\tilde{\sigma} - T u_3^*\tilde{\sigma}}{2 R_0}\right)ds,
\end{aligned}
\]
where the term multiplying $\tilde{\sigma}'$ under the integral sign disappears due to the boundary conditions $\sigma(0) = \sigma(L) = 0$. The term $\Psi_1$ corresponds to the first variation of Eq.~\eqref{eq:energy_no_rotation_disc} about the reference configuration, namely $\Psi_1[\tilde{\sigma}, \, p_0] = \delta\Psi(0,\,p_0)[R_0\tilde{\sigma},\,p_1]$. This is zero for all the admissible $\tilde{\sigma}$ if and only if
\begin{equation}
\label{eq:p0}
p_0 = T u_3^*,
\end{equation}
so that the reference configuration is in equilibrium also when $u_3^*\neq0$.

\begin{figure}[b!]
\centering
\includegraphics[align=c, height=0.27\textwidth]{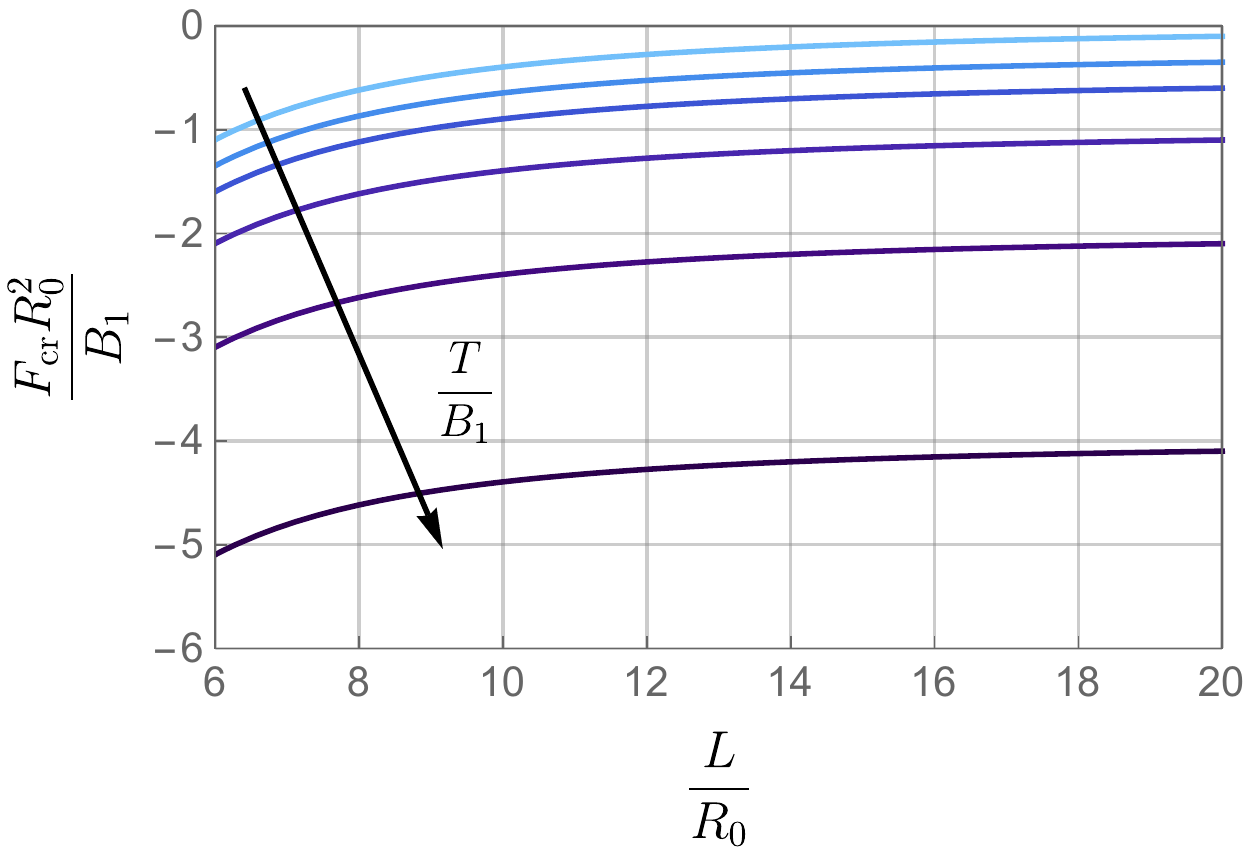}
\includegraphics[align=c, height=0.27\textwidth]{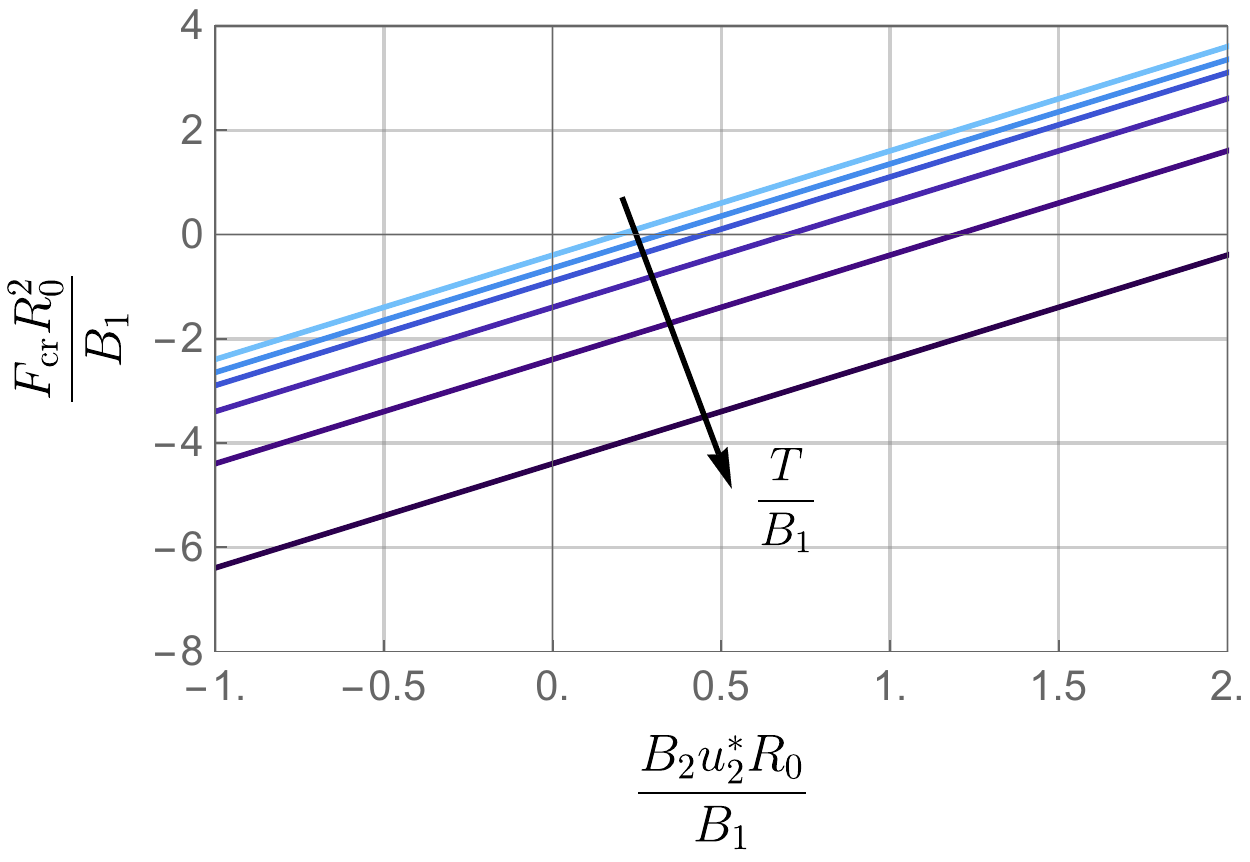}
\includegraphics[align=c, height=0.4\textwidth]{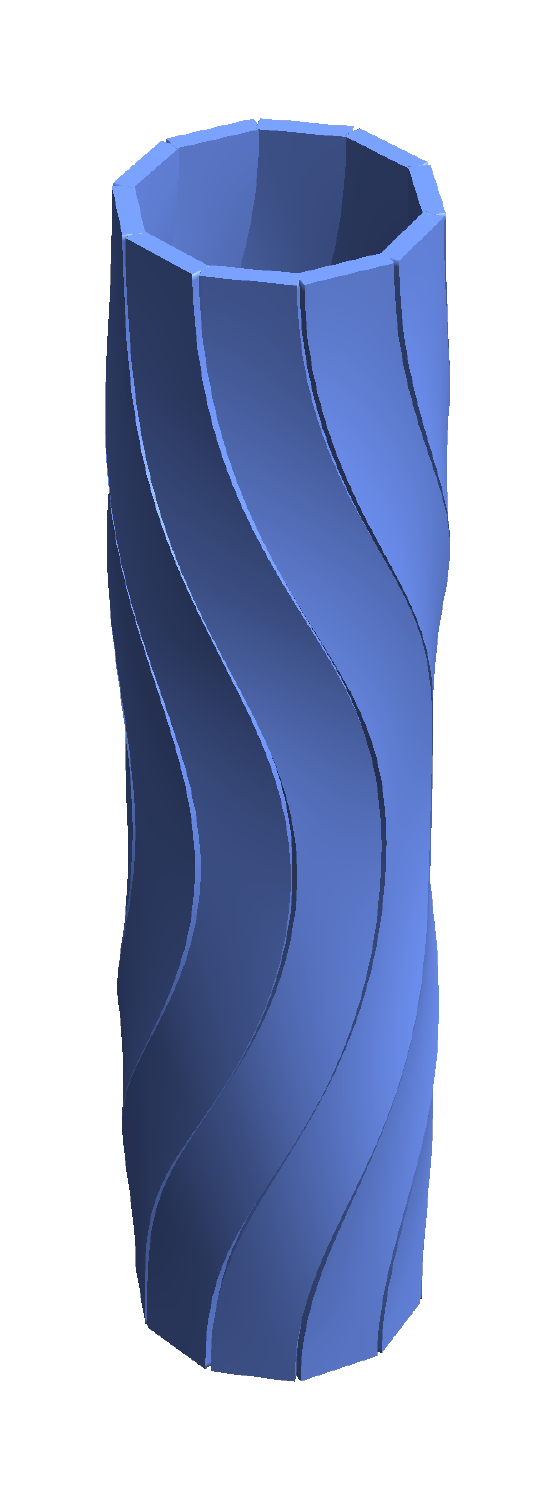}
\caption{Non-dimensional critical buckling load $F_\text{cr}R_0^2/B_1$ versus $L/R_0$ when $u_2^*=0$ (left) and $F_\text{cr}R_0^2/B_1$ versus $B_2 u_2^*R_0/B_1$  when $L/R_0=10$ (center) for $T/B_1 = 0,\,1/4,\,1/2,\,1,\,2,\,4$. The arrows denote the direction in which $T/B_1$ grows. (Right)
bifurcation mode for $n=10$ and $L/R_0 = 10$. The amplitude of $\sigma$ has been set to $0.25 R_0$ for the sake of graphical illustration.}
\label{fig:clamped_double}
\end{figure}

Finally, $\Psi_2$ is a functional of $\tilde{\sigma}$ and $p_1$. In particular, its expression reads
\[
\begin{aligned}
\Psi_2[\tilde\sigma, \, p_1] & = \frac{1}{4 R_0^2}\int_0^L\left[2 R_0 \tilde{\sigma} \left(R_0^2 \tilde{\sigma}'' \left(B_{1} u_1^* \cot \left(\frac{\pi }{n}\right)+B_{2} u_2^* \left(\csc ^2\left(\frac{\pi }{n}\right)+1\right)\right)+2 p_1 \cot \left(\frac{\pi }{n}\right)\right)+\right.\\
& + R_0^2 \tilde{\sigma}'^2 \left(\csc ^2\left(\frac{\pi }{n}\right) (B_{1}+2 B_{2} R_0 u_2^*)+2 B_{1} R_0 u_1^* \cot \left(\frac{\pi }{n}\right)-B_{1}+2 B_{2} R_0 u_2^*\right)+\\
& \left. + \cot ^2\left(\frac{\pi }{n}\right) \tilde{\sigma}^2 \left(-2 B_{2} R_0 u_2^*+F R_0^2+T\right)\right] ds.
\end{aligned}
\]
To study the stability of the reference configuration, we adopt again the Trefftz criterion and compute the first variation of $\Psi_2$ with respect to $\tilde{\sigma}$ and $p_1$ \citep{Ba_ant_2010}. The natural boundary condition is identically satisfied, as in case (a), owing the edge compatibility condition. The Euler-Lagrange equations now read
\begin{equation}
\label{eq:eul_lag_const_rot}
\left\{
\begin{aligned}
&\cot \left(\frac{\pi }{n}\right) \left(B_{1} R_0^2 \tilde{\sigma}''+\tilde{\sigma} \left(2 B_{2} R_0 u_2^*-F R_0^2-T\right)\right)=2 p_1 R_0,\\
&\int_0^L\tilde{\sigma}\,ds = 0.
\end{aligned}
\right.
\end{equation}

We seek for solutions of the form
\[
\tilde{\sigma}_m(s) = \sin\left(\frac{m\pi}{L}s\right),\qquad m\in\mathbb{N}\setminus\{0\}.
\]
It is easily verified that the mode $\tilde{\sigma}_m$ satisfies the second equation of Eq.~\eqref{eq:eul_lag_const_rot} if and only if $m$ is even. Using the first equation, we get that the load at which $\tilde{\sigma}_m$ is a solution of the linear problem is given by
\[
F = F_m = -\frac{\pi ^2 B_1 m^2}{L^2}-\frac{T-2 B_2 R_0 u_2^*}{R_0^2}\qquad\text{with}\qquad p_1 = 0.
\]
We  conclude that the critical mode is $m = 2$ and the critical load is given by $F_\text{cr} = F_2$. The configuration corresponding to this critical mode is reported in Fig.~\ref{fig:clamped_double} (right). In the figure (left and center) we also plot the non-dimensional critical load, exploring the influence of the torsional stiffness and of the natural curvature $u_2^*$. The effect of these quantities on the critical load is similar to what is reported for the previous case. Interestingly, now the first term in the critical load coincides with the buckling load of a doubly clamped beam  that buckles by bending along the $\vect{d}_1^k$ direction, in agreement with the fact that, by constraining the relative rotation of the ends, the rods extremities cannot undergo a lateral displacement.

\subsection{Infinite number of rods: Linear and post-buckling analysis}

We examine now the buckling of a rod assembly subjected to an axial load in the limit of an infinite number of rods, so that the potential elastic energy is only a function of $\gamma$, recall Eq.~\eqref{eq:energy_gamma}. We adopt the asymptotic expansion of \eqref{eq:asympt_exp_n_inf}, which led to Eq.~\eqref{eq:rtz_n_infinity} and  \eqref{eq:u_n_inf}, and neglect higher order terms. For the sake of simplicity, since from Eq.~\eqref{eq:u_n_inf} $u_1 = \gamma'$, in this section we also adopt the thin rod approximation, which results in greatly simplified expressions.

\subsubsection{Case (a): Free rotation}
\label{sec:lin_stab_free}
We assume that a force $F\vect{e}_z$ is applied to each rod at $s = L$. The work done on a single rod is given by
\[
F (z - L) = F \left(\int_0^L\sqrt{\frac{1-R_0^2 \gamma ^2 \gamma '^2}{\gamma ^2+1}}\,ds - L\right).
\]
so that, the total energy of a single rod reads
\begin{equation}
\label{eq:energy_axial_force}
\Psi[\gamma] = \frac{B_1}{2} \int_0^L (u_1 - u_1^*)^2\,ds - F \left(\int_0^L\sqrt{\frac{1-R_0^2 \gamma ^2 \gamma '^2}{\gamma ^2+1}}\,ds - L\right).
\end{equation}
Thus, we can compute the non-linear Euler-Lagrange equation arising from the functional \eqref{eq:energy_axial_force} so that
\[
F \gamma  z'_\infty \left(R_0^2 \left(\left(\gamma ^3+\gamma \right) \gamma ''+\left(2 \gamma ^2+1\right) \gamma '^2\right)-1\right)+B_1 \left(\gamma ^2+1\right) \gamma '' \left(R_0^2 \gamma ^2 \gamma '^2-1\right)^2 = 0,
\]
where $z'_\infty$ is given by Eq.~\eqref{eq:rtz_n_infinity}. We observe that $\gamma(s) = 0$ is a solution. To study the stability of the straight cylindrical configuration, we linearize the system about the relaxed configuration, obtaining
\[
B_1\gamma'' = F \gamma.
\]

\begin{figure}[t!]
\centering
\includegraphics[height=0.4\textwidth]{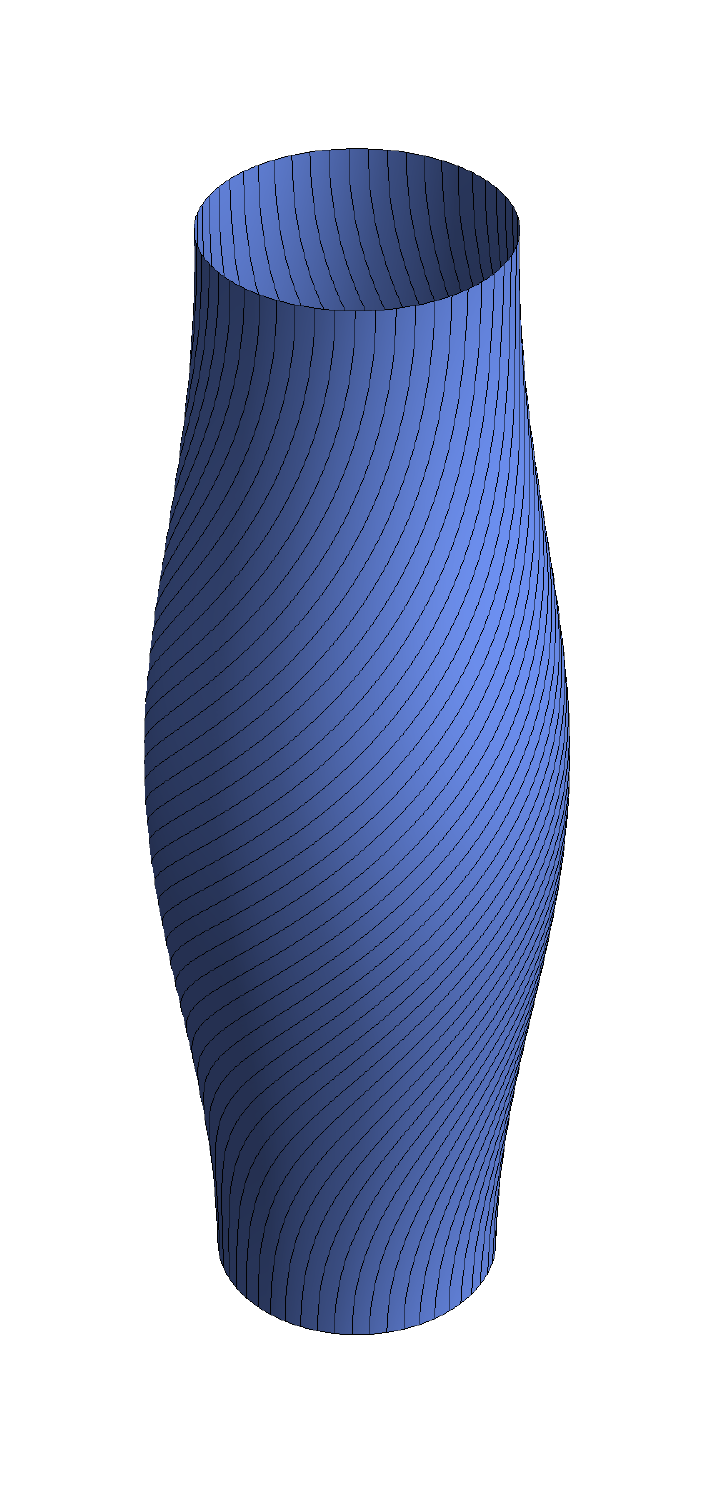}\qquad \includegraphics[height=0.4\textwidth]{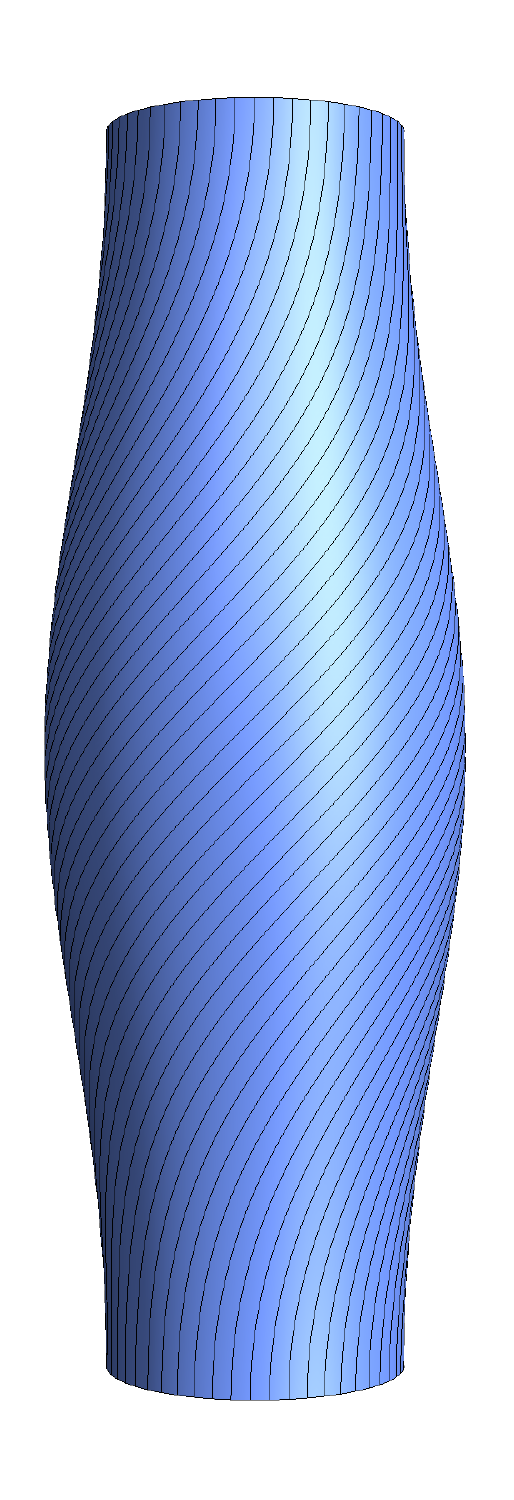}
\caption{Critical buckling mode when $L/R_0 = 10$. The amplitude of the mode $A$ as been set equal to $1$ for the sake of graphical clarity.}
\label{fig:crit_mode_rot}
\end{figure}

To judge the linear stability of the trivial solution $\gamma = 0$, let
\[
\gamma_m(s) = A \sin\left(\frac{m \pi s}{L}\right),\qquad m\in\mathbb{N}\setminus\{0\} .
\]
The function $\gamma_m$ is a solution of the linearized problem whenever
\begin{equation}
\label{eq:marg_stab_n_inf}
F = F_m = -\frac{\pi ^2 m^2 B_1}{L^2} ,
\end{equation}
thus, the critical buckling load $F_\text{cr}$ is obtained by setting $m = 1$.
The buckled configuration exhibits a single bulge whose chirality depends on the sign of $\gamma$, as shown in Fig.~\ref{fig:crit_mode_rot}. Having neglected $B_2$ and $T$, the critical load coincides with the first term in Eq.~(\ref{Fcr_a}), and hence with the critical load of an isolated rod that buckles by bending along the $\vect{d}_1^k$ direction with constrained rotation at both ends but free lateral translation. We note that here $B_2 =0$ and thus the collective constraints prevent the rods from buckling along their most compliant direction. Thus, given $B_1$, the critical buckling load predicted by the theory in the limit $n\rightarrow \infty$ is exact and in fact independent of $n$. Furthermore, this result shows that the mechanical effect of the assembly constraints on the critical buckling load is minor in the thin rod approximation. We shall see later that this effect manifests itself in the post-buckling regime.

\subsubsection{Case (b): Constrained rotation}

We further constrain the system by forbidding the relative rotation of the two cylinder ends. Namely, we introduce the constraint
\begin{equation}
\label{eq:thetaninf}
\theta(0) = \theta(L).
\end{equation}
Exploiting Eq.~\eqref{eq:rtz_n_infinity}, we can express Eq.~\eqref{eq:thetaninf} as a constraint on $\gamma$, enforced through the enforcement of a scalar Lagrange multiplier $p$. Hence, the Lagrangian of a single rod becomes
\begin{equation}
\label{eq:energy_no_rotation}
\Psi[\gamma,\,p]= \frac{B_1}{2} \int_0^L (\gamma' - u_1^*)^2\,ds - F \left( \int_0^L\sqrt{\frac{1-R_0^2 \gamma ^2 \gamma '^2}{\gamma ^2+1}}\,ds - L\right) + p \int_0^L\frac{\gamma}{R_0(\gamma^2+1)}\,ds.
\end{equation}
As before, we set $\gamma(0) = \gamma(L) = 0$, so that the Euler-Lagrange equations relative to \eqref{eq:energy_no_rotation} read
\begin{small}
\begin{equation}
\label{eq:non_lin_gamma_p}
\left\{
\begin{aligned}
&\frac{\left(\gamma ^2+1\right) \gamma '' \left(F R_0^2 \gamma ^2-B_1 \left(\gamma ^2+1\right) \left(R_0^2 \gamma ^2 \gamma '^2-1\right) z'_\infty\right)+F R_0^2 \left(2 \gamma ^3+\gamma \right) \gamma '^2-F \gamma }{\left(\left(\gamma ^2+1\right) \left(1-R_0^2 \gamma ^2 \gamma '^2\right)\right)^{3/2}}=\frac{p \left(1-\gamma ^2\right)}{R_0 \left(\gamma ^2+1\right)^2},\\
&\int_0^L\frac{\gamma}{R_0(\gamma^2+1)}\,ds=0.
\end{aligned}
\right.
\end{equation}
\end{small}
We observe that $(\gamma(s) = 0,\,p=0)$ is a solution for the system above. Following the same steps as in the previous section, the linearized Euler-Lagrange equations about the reference configuration read
\[
\left\{
\begin{aligned}
&B_1 \gamma'' - F \gamma - \frac{p}{R_0}=0,\\
&\int_0^L\gamma\,ds = 0 ,
\end{aligned}
\right.
\]
where the latter equation arises from the first variation with respect to the Lagrange multiplier.
\label{sec:lin_stab_constrained}
\begin{figure}[t!]
\centering
\includegraphics[height=0.4\textwidth]{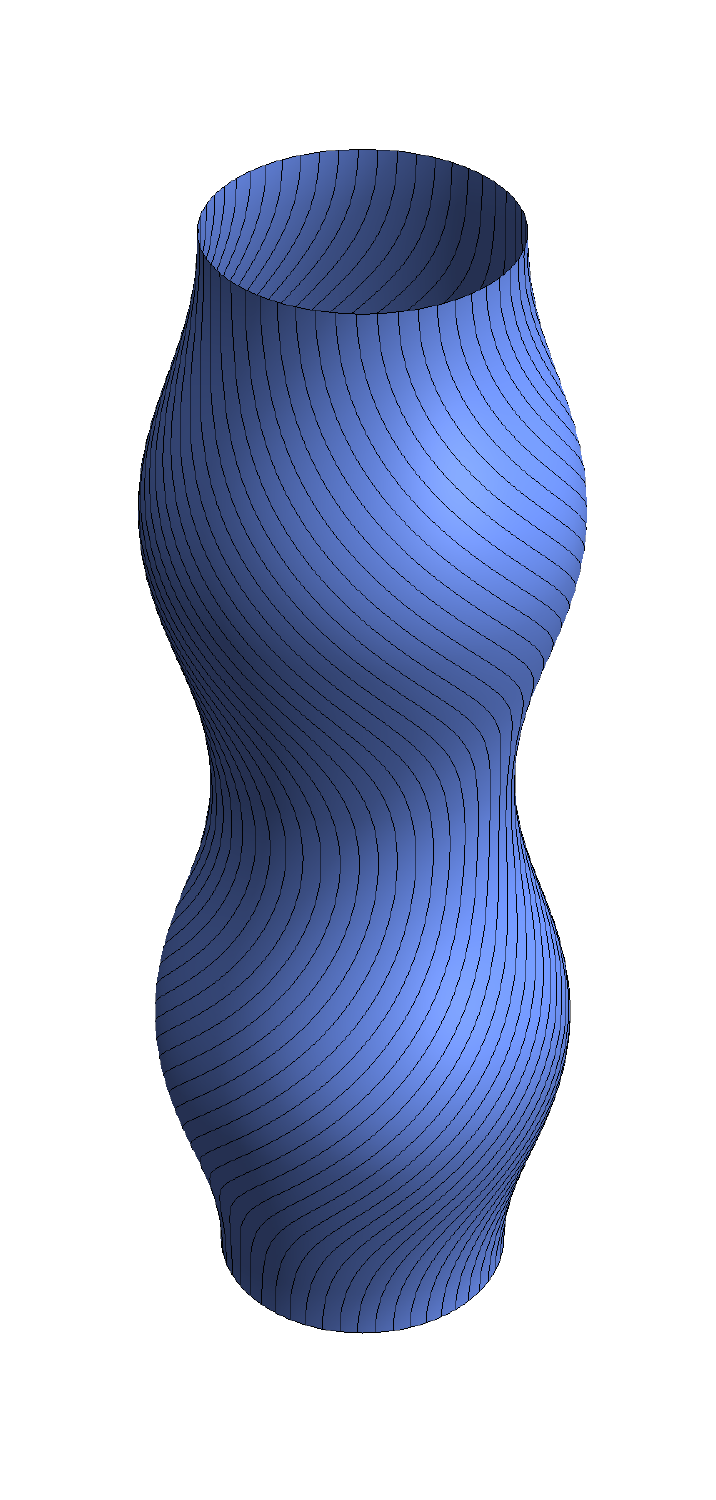}\qquad\includegraphics[height=0.4\textwidth]{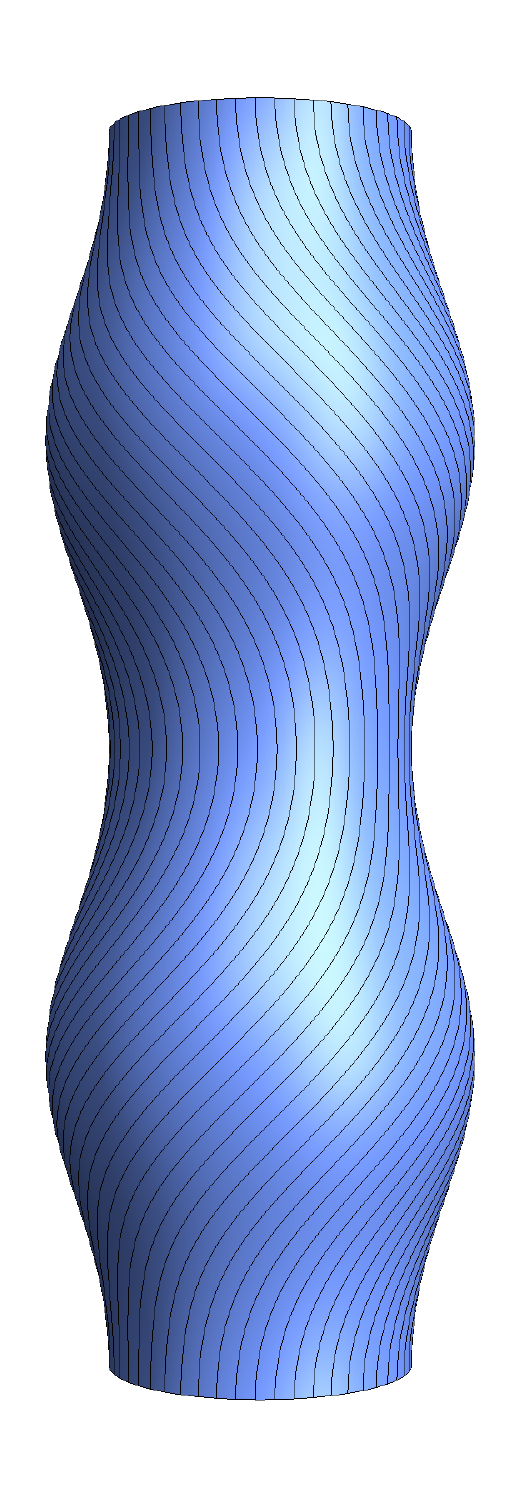}
\caption{Critical buckling mode when $L/R_0 = 10$. The amplitude of the mode $A$ as been set equal to $1$ for the sake of graphical clarity.}
\label{fig:no_rotation_crit_mode}
\end{figure}
To study the linear stability of the reference configuration, we consider the functions
\[
\gamma_m(s) = A \sin\left(\frac{m \pi s}{L}\right), \qquad m\in\mathbb{N}\setminus\{0\},
\]
which are solutions of the linearized problem whenever $m$ is even and
\begin{equation}
\label{eq:marg_stab_n_inf_no_rot}
F = F_m = -\frac{\pi ^2 B_1 m^2}{L^2} \quad \text{and} \quad p = 0.
\end{equation}
We observe that the Lagrange multiplier is always null, indicating that the reaction exerted by the constraint is zero. The effect of the constraint is the suppression of the odd wavenumbers. The critical mode exhibits two bulges possessing opposite chiralities, as shown in Fig.~\ref{fig:no_rotation_crit_mode}. Again, the critical load coincides with that of the finite collection of rods with $B_2=T=0$ and with that of a doubly clamped isolated rod buckling along the $\vect{d}_1^k$ direction.

\subsubsection{Numerical post-buckling analysis}
\begin{figure}[t!]
\centering
\subfloat[]{\includegraphics[width=0.5\textwidth]{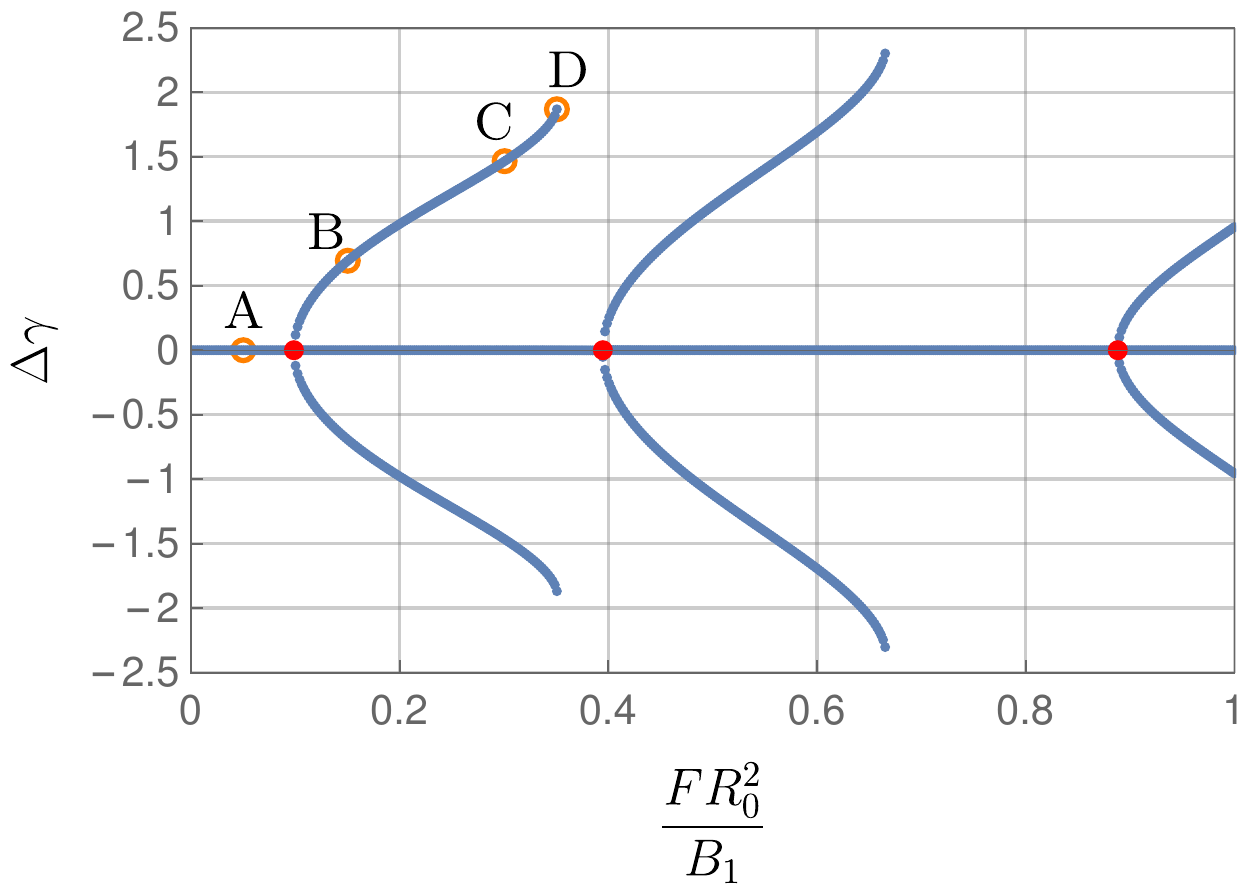}\label{fig:bif_diag_free}}\subfloat[]{\includegraphics[width=0.5\textwidth]{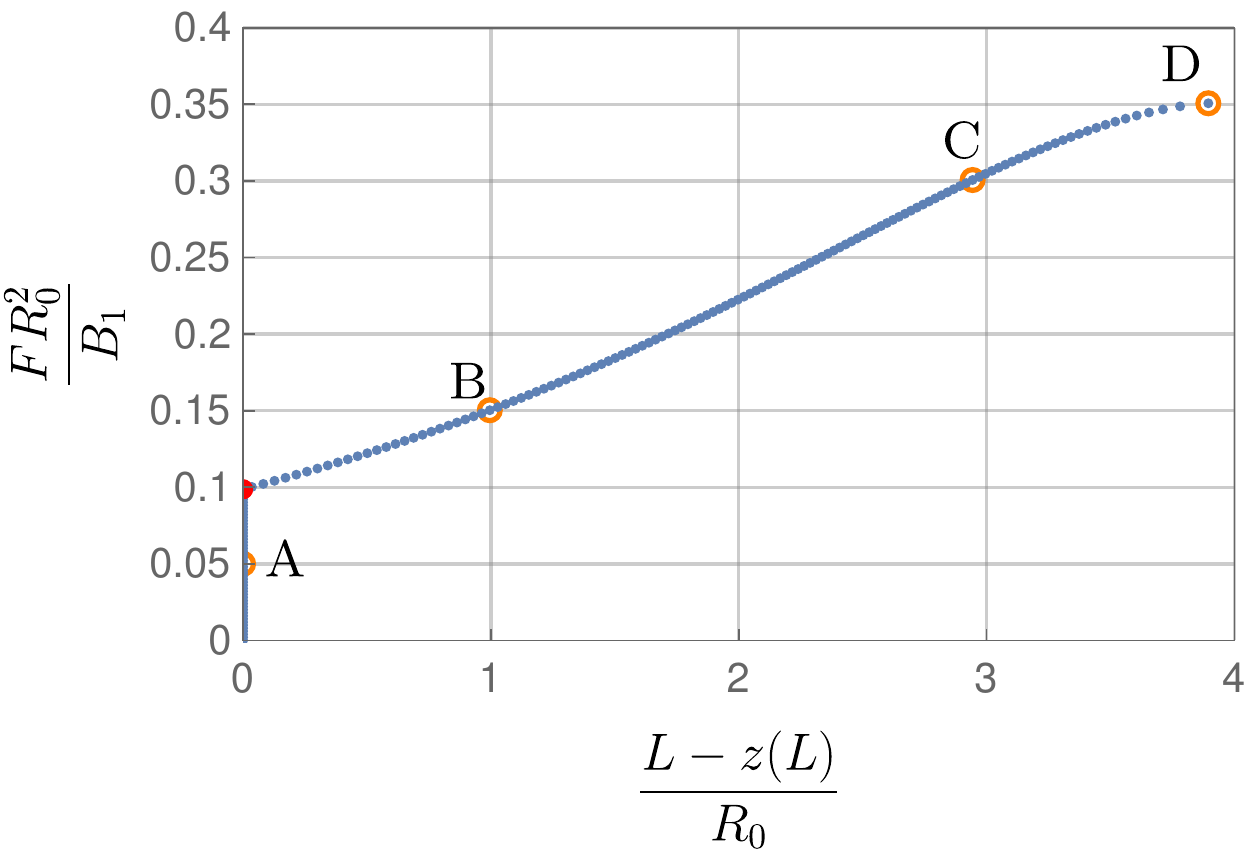}\label{fig:forza_spos_free}}\\
\subfloat[]{\includegraphics[width=0.5\textwidth]{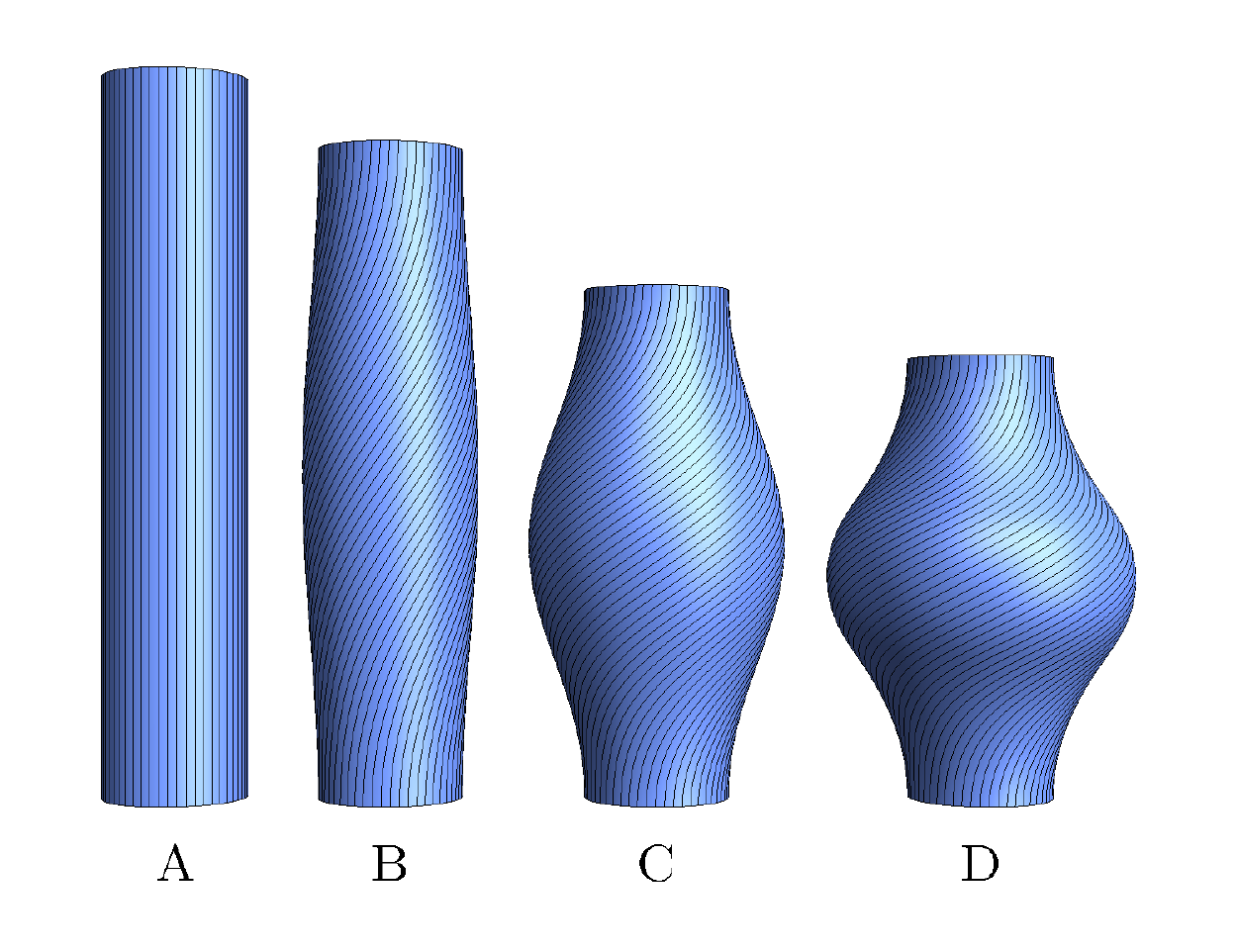}\label{fig:configurations_free}}\subfloat[]{\includegraphics[width=0.5\textwidth]{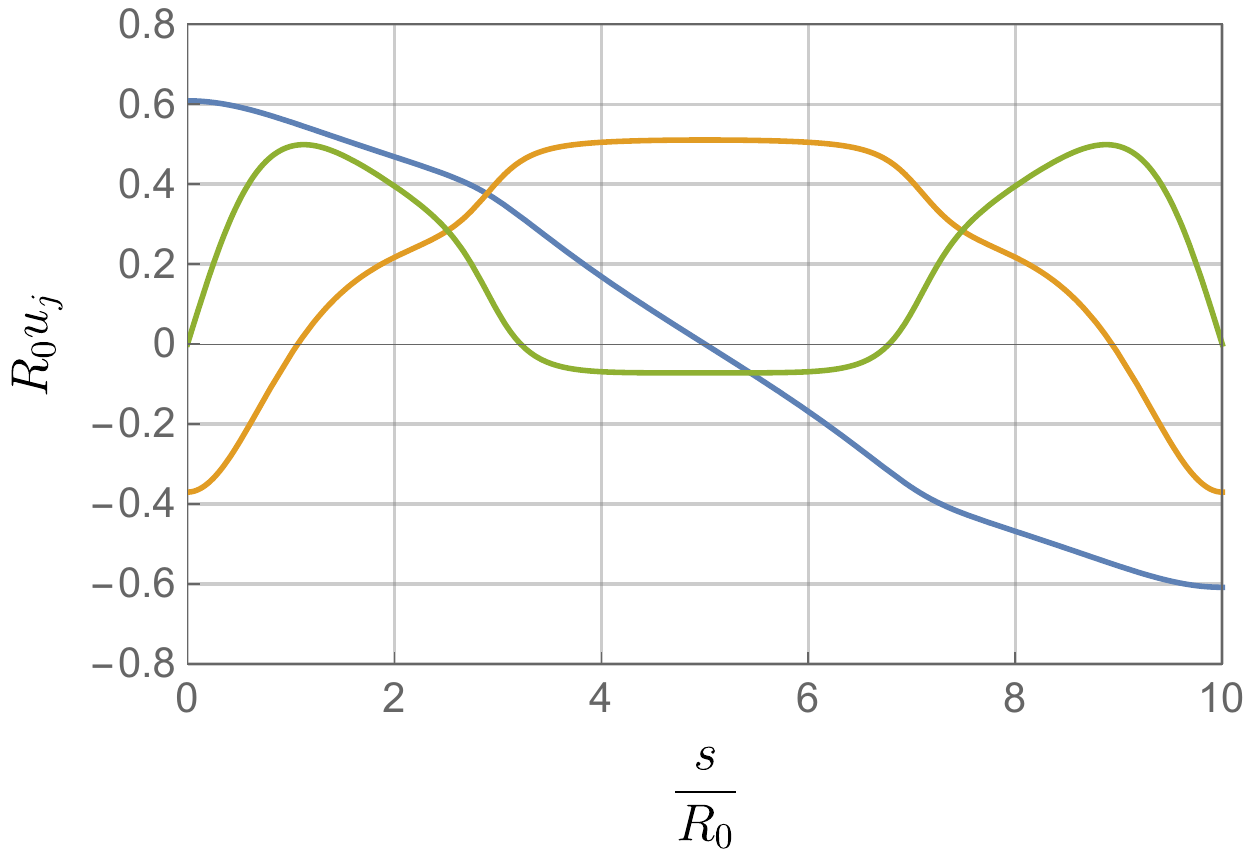}\label{fig:strain_free}}
\caption{(a) Bifurcation diagram $FR_0^2/B_1$ versus $\Delta \gamma$ when $L/R_0 = 10$. The red dots denote the theoretical marginal stability threshold arising from the linear theory, see Eq.~\eqref{eq:marg_stab_n_inf}. (b) Plot of the non-dimensionalized applied force $FR_0^2/B_1$ versus vertical displacement of the assembly $L - z(L)$ non-dimensionalized  with respect to $R_0$ for the branch corresponding to the critical mode. (c) Buckled configuration for $L=10$ and $FR_0^2/B_1 = 0.1,\,0.15,\,0.3$ and $0.35$. (d) Plot of $u_1$, $u_2$ and $u_3$ (blue, orange and green line respectively) as a function of the arclength $s$. We observe that all the components of the strain vector $\vect{u}$ are bounded.}
\end{figure}
To study the post-buckling behavior of the rod assembly in the presence of an axial load in the limit case of $n\rightarrow\infty$, we have implemented a numerical algorithm to solve the fully non-linear problem arising from the minimization of the functionals in Eqs.~\eqref{eq:energy_axial_force} and \eqref{eq:energy_no_rotation} in the absence of natural curvature, {\it i.e.} for $u_1^*=0$. The numerical procedure is based on the finite element method using the software FEniCS \citep{LoggMardalEtAl2012a}. To compute the bifurcation diagram of the problem, we use the deflated continuation method exploiting the library Defcon, see \citep{Farrell_2015,farrell2016computation} for details of the continuation algorithm. The control parameter of the problem is provided by the force $F$ applied to each rod and it is sequentially incremented by a constant $\delta F$ when a solution is found.

All the physical quantities of the system are non-dimensionalized with respect to $B_1$ and $R_0$. The domain $(0,\,L)$ is discretized by dividing it into 1000 equal subintervals. The local shear $\gamma$ is discretized using continuous, piece-wise linear functions while the Lagrange multiplier $p$ is treated as an additional degree of freedom of the problem.

We report next the results of the numerical simulations when we allow a relative rotation of the two ends of the cylinder and when both ends are kept fixed.
\paragraph{Case (a): free rotation}
\begin{figure}[t!]
\centering
\subfloat[]{\includegraphics[width=0.5\textwidth]{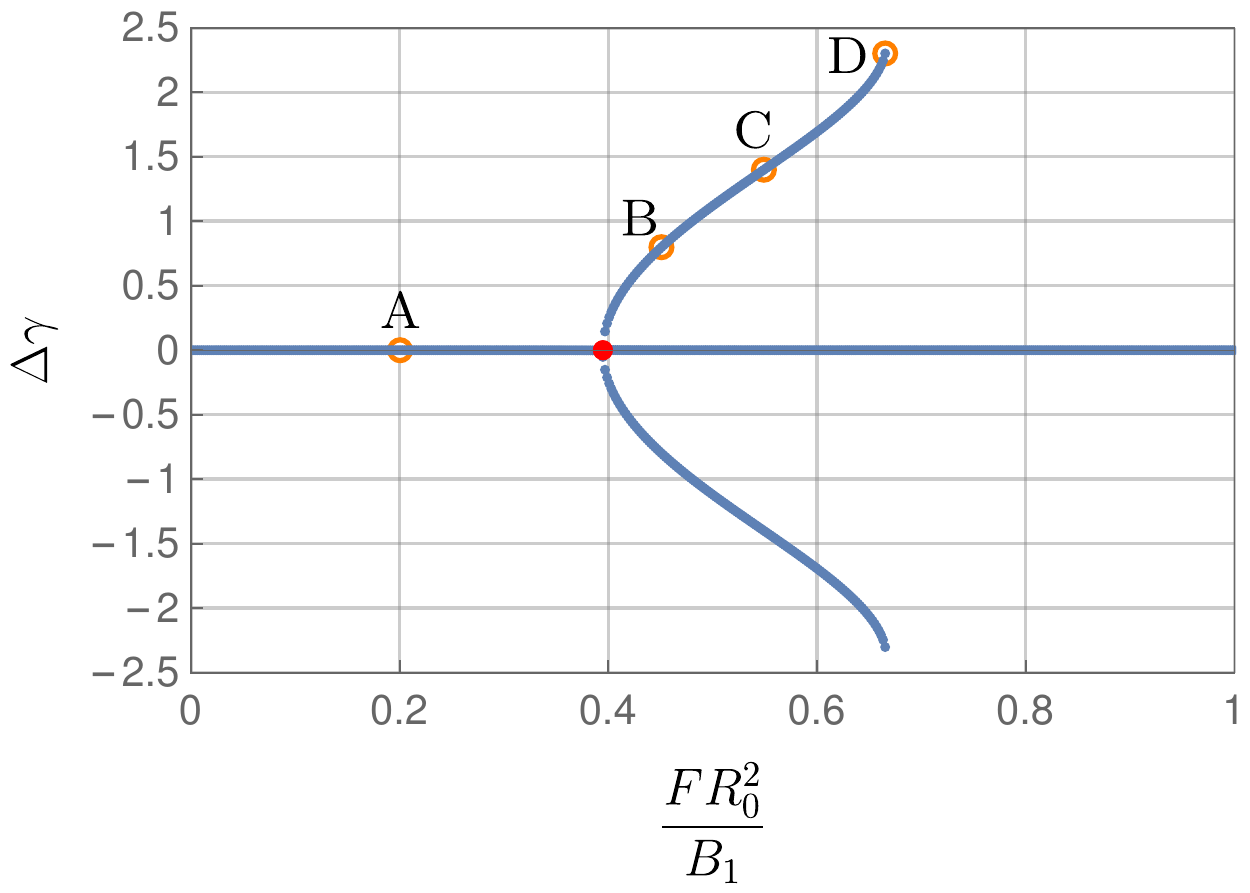}\label{fig:bif_diag_constrained}}\subfloat[]{\includegraphics[width=0.5\textwidth]{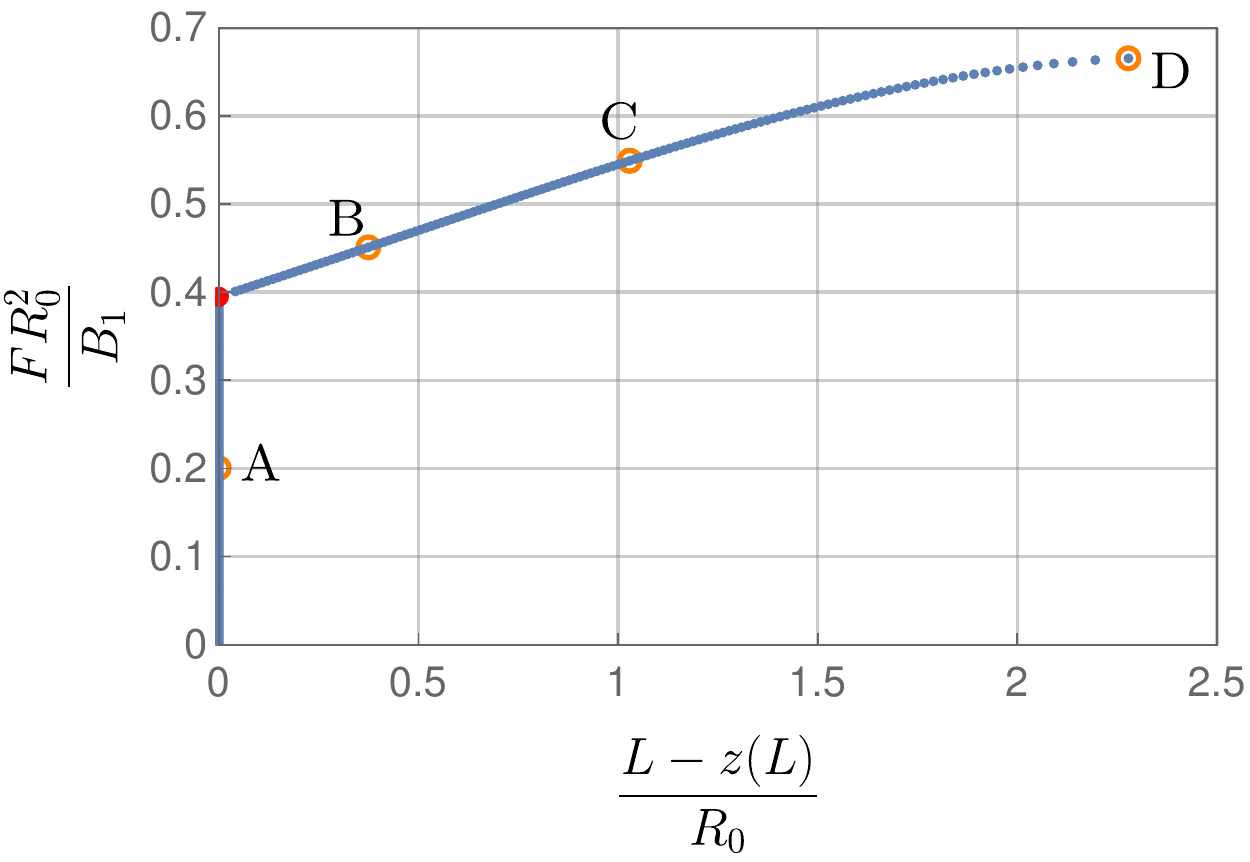}\label{fig:forza_spos_constrained}}\\
\subfloat[]{\includegraphics[width=0.5\textwidth]{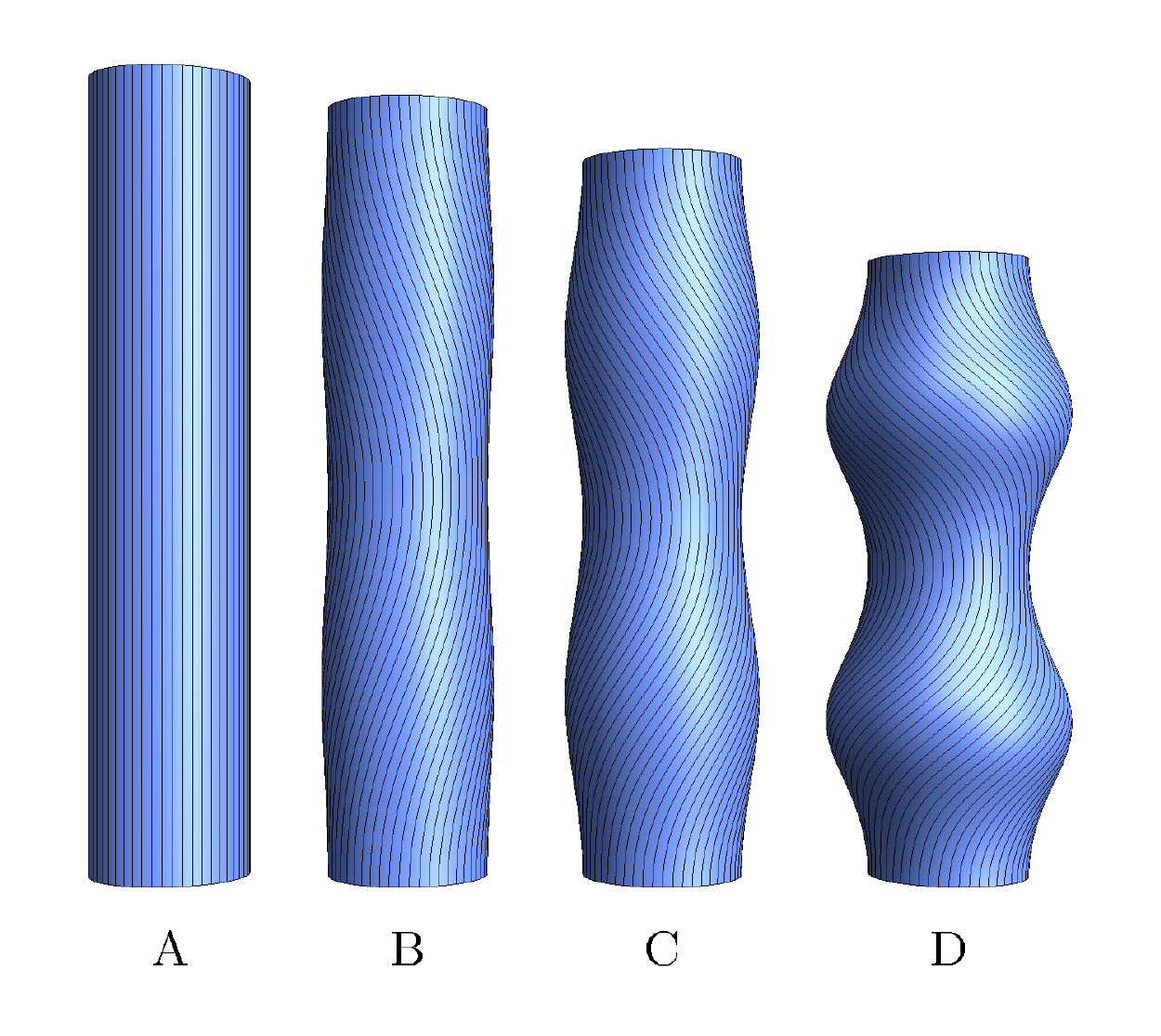}\label{fig:configurations_constrained}}\subfloat[]{\includegraphics[width=0.5\textwidth]{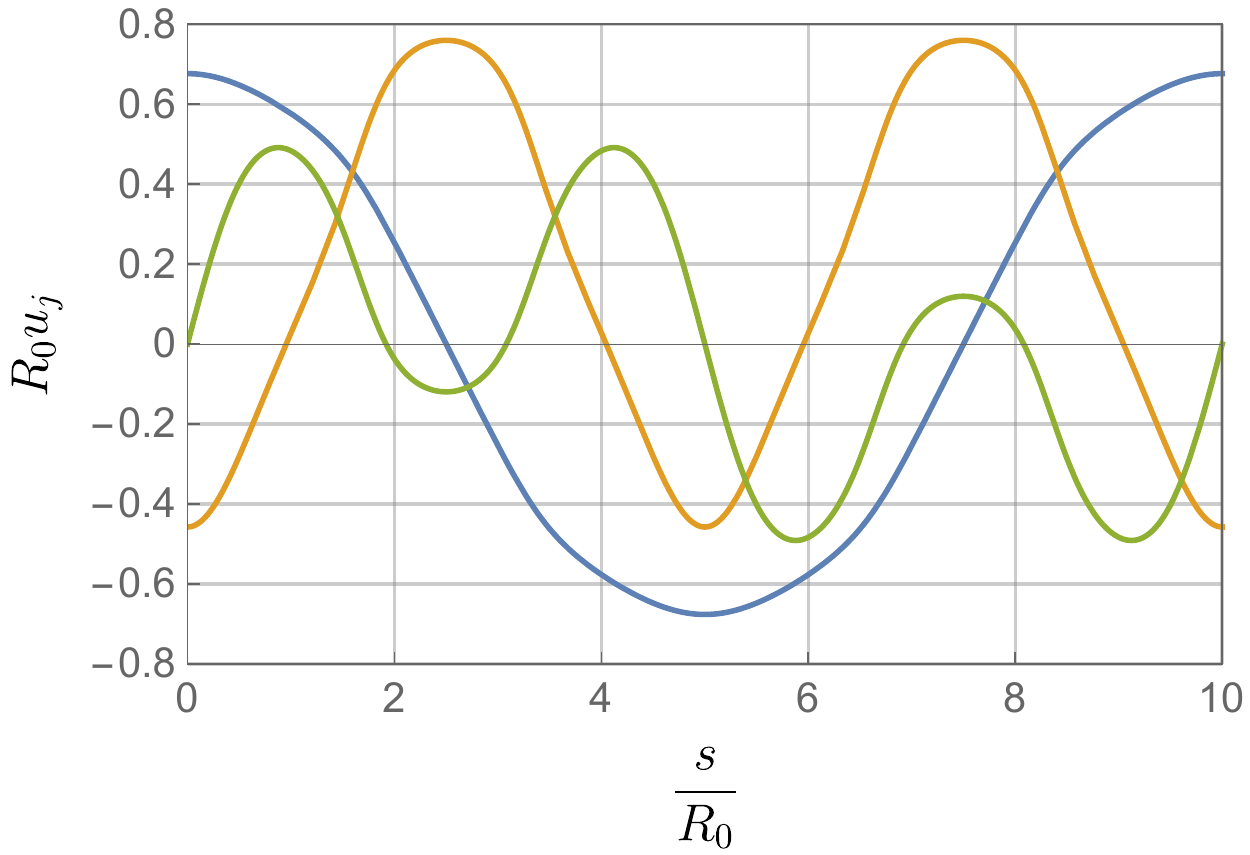}\label{fig:strain_constrained}}
\caption{(a) Bifurcation diagram $\Delta \gamma$ versus $FR_0^2/B_1$ when $L/R_0 = 10$ and $\theta(0)=\theta(L)=0$. The red dots denote the theoretical marginal stability threshold arising from the linear theory, see Eq.~\eqref{eq:marg_stab_n_inf_no_rot}. (b) Plot of the non-dimensionalized applied force $FR_0^2/B_1$ versus vertical displacement of the assembly $L - z(L)$ non-dimensionalized  with respect to $R_0$ for the branch corresponding to the critical mode. (c) Buckled configuration for $L=10$ and $FR_0^2/B_1 = 0.2,\,0.45,\,0.55$ and $0.67$. (d) Plot of $u_1$, $u_2$ and $u_3$ (blue, orange and green line respectively) as a function of the arclength $s$. We observe that all the components of the strain vector $\vect{u}$ are bounded.}
\end{figure}
Let us introduce the quantity
\[
\Delta\gamma = \sgn(\gamma'(0))\left(\max_{s\in(0,L)}\gamma(s) - \min_{s\in(0,L)}\gamma(s)\right).
\]
We report in Fig.~\ref{fig:bif_diag_free} the bifurcation diagram resulting from the numerical simulation when we leave the top basis free to rotate. We have verified that the bifurcated branches have their origins in correspondence of the theoretical critical load computed in Section~\ref{sec:lin_stab_free}, validating our numerical code.
We further observe that all the bifurcations have the typical shape of a supercritical pitchfork. The force-displacement curve is also reported in Fig.~\ref{fig:forza_spos_free} in terms of non-dimensional quantities. The post-buckling evolution of the assembly is depicted in Fig.~\ref{fig:configurations_free}, and exhibits a bulge at the middle of the rod assembly whose radius increases as we increment the applied force.
We remark that all the components of the strain vector $\vect{u}$ are bounded, as shown in Fig.~\ref{fig:strain_free}. This shows that the terms
\begin{equation}
\label{eq:neglected_terms}
B_2(u_2-u_2^*)^2 , \qquad T(u_3 - u_3^*)^2,
\end{equation}
in the potential energy \eqref{eq:elastic_energy} can in fact be neglected: even though $B_2$ and $T$ are negligible when the thin rod approximation applies, bending, $u_2$, and torsion, $u_3$, curvatures could blow up. However, this does not happen even in the post-buckling regime.

\paragraph{Case (b): constrained rotation}

The introduction of the constraint $\theta(0) = \theta(L) = 0$ requires the introduction of a Lagrange multiplier $p$, as discussed in Section~\ref{sec:lin_stab_constrained}. This modification of the problem affects the bifurcation diagram as shown in Fig.~\ref{fig:bif_diag_constrained}: the bifurcated branches corresponding to the wavenumber $m = 1$ and $m = 3$ disappear and the only bifurcated branch corresponds to the critical wavenumber $m = 2$. As well as in the unconstrained case shown in the previous section, the bifurcation diagram exhibit all the characteristics of a supercritical pitchfork bifurcation. As for the previous analysis, the non-dimensional force-displacement curve is reported in Fig.~\ref{fig:forza_spos_constrained}.
The non-linear evolution of the buckled assembly exhibits two bulges placed one on the top of the other, as revealed by the linear stability analysis of Section~\ref{sec:lin_stab_constrained}, see also Fig.~\ref{fig:configurations_constrained}. We notice that the rotation induced by the formation of the two bulges is balanced by their opposite chiralities. As before, all the components of the strain vector are bounded, as shown in Fig.~\ref{fig:strain_constrained}, confirming the validity of the thin rod approximation.

\section{Discussions of the results and concluding remarks}

In this paper, we have studied the mechanics of assemblies composed of $n$ identical rods where each rod is interlocked to both the adjacent ones. We have developed the mathematical model of this structure, as inspired by the euglenoid pellicle, using the special Cosserat theory of rods.
We have characterized the mechanics of axisymmetric sheets composed of sliding elastic beams, investigating  buckling in presence of an axial load. 
As compared to a continuous equivalent shell, the sliding degrees of freedom make our structures much more compliant. Yet, the sliding constraints between the rods introduce non-trivial and highly nonlinear couplings between compression and twist of the assembly, and between the different deformation modes of the rods. One of the main results of our paper is summarized in Eq.~(\ref{Fcr_a}), which shows how the constraints between rods in our assembly modify the bare buckling load of a single rod. It shows how the helical shapes of rods induced by the constraints mobilize
the higher bending modulus ($B_1$ rather than $B_2$, which is much smaller in the case of thin rods)  and the 
 torsional stiffness $T$, and hence increase the buckling load. Moreover, it shows
how spontaneous curvature in the non-buckling direction can either increase or decrease the critical buckling load.

Our work extends the study of \cite{Noselli_2019}, where deformations of the rod assembly were assumed to remain cylindrical, but is limited to  axisymmetry. The axisymmetric shapes reported here are very similar to those observed as Euglena cells move in a fluid or in a confined environment, where they tune their shape and develop bulges  \citep{buetow1968biology, Noselli2019a}. The structure of such bulges is very similar to the critical modes predicted by our stability analysis. Thus, our work is pertinent to the study of microswimmers and the design of bioinspired microrobots. The assumption of axisymmetry can be too restrictive in other cases. Euglena cells are also capable of exhibiting non-axisymmetric deformations, and numerical evidence suggests that cylindrical assemblies of sliding rods can produce non-axisymmetric shapes \citep{Arroyo_2014}. Going back to our study, even though the homogeneous boundary conditions on the sliding $\sigma$ favor axisymmetric buckling modes, it is possible that non-axisymmetric buckling modes become critical for large aspect ratio $L/R_0$.

Future efforts will be devoted to the weakening of the constraint given by Eq.~\eqref{eq:compatibility}, allowing us to enforce the compatibility only in a portion of the beam. We will also study how to program these sheets to obtain target configurations by controlling the natural curvatures and torsion of the rods. Furthermore, it would be important to develop computational methods to approximate the fully non-linear problem also in the discrete case, as well as to obtain an experimental validation of the proposed model through the realization of these structures by additive manufacturing techniques.

\section*{Acknowledgments}
ADS, GN, and DR acknowledge the support of the European Research Council (AdG-340685-MicroMotility). MA acknowledges the support of the European Research Council (CoG-681434), the Generalitat de Catalunya (2017-SGR-1278 and ICREA Academia prize for excellence in research). 

\appendix
\section{Cylindrical configurations}
\label{app:cylinders}
In this Appendix, we show how to recover the mathematical description developed in \cite{Noselli_2019} by restricting the kinematics to cylindrical configurations. Let
\begin{equation}
\label{eq:restr_cin_cyl}
\left\{
\begin{aligned}
&\rho(s) = R,\\
&\theta(s) = \beta s,\\
&\alpha(s) = 0,
\end{aligned}
\right.\qquad \beta>0,\,R>0.
\end{equation}
Enforcing the inextensibility constraint of Eq.~\eqref{eq:inex_rho}, we get the following expression of the actual height $z$
\begin{equation}
\label{eq:zcyl}
z(s)=\pm s \sqrt{1-\beta ^2 R^2}.
\end{equation}
Without loss of generality, we can assume that $z>0$. Furthermore, since the coordinate $z$ given by Eq.~\eqref{eq:zcyl} must be a real number, we get $\beta R < 1$.
By projecting the compatibility constraint \eqref{eq:compatibility} along the $z$ direction we obtain
\[
\sigma = \frac{\beta  h R}{\sqrt{1-\beta ^2 R^2}},
\]
while the radial and the azimuthal components lead to
\begin{equation}
\label{eq:Rcyl}
R = \frac{1}{2} h \sqrt{1-\beta ^2 R^2} \cot \left(\frac{\pi }{n}-\frac{\beta ^2 h R}{2 \sqrt{1-\beta ^2 R^2}}\right).
\end{equation}
Defining $\sin\xi = \beta R$, we can rewrite Eq.~\eqref{eq:Rcyl} so that 
\[
R = \frac{h}{2} \cot \left(\frac{\pi }{n}-\frac{h \sin\xi\tan\xi}{2 R}\right)\cos\xi.
\]
This last equation is exactly Eq.~(10) as reported in \cite{Noselli_2019}.

\section{Third order expansion}
\label{app:third_order}
Since both $\alpha_1$ and $\alpha_2$ in Eq.~\eqref{eq:second_order_expansion} are zero, one could ask if $\alpha$ itself is null for all the possible axisymmetric configurations satisfying the compatibility equation \eqref{eq:compatibility}. The answer to this question is negative. Indeed, by computing the third order expansion in $\varepsilon$ of $\rho,\,\theta,\,z$ and $\alpha$, {\it i.e.} by writing
\[
\left\{
\begin{aligned}
\rho(s) &= R_0 + \varepsilon \rho_1(s)+\varepsilon^2\rho_2(s)+\varepsilon^3\rho_3(s)+o(\epsilon^3),\\
\theta(s) &= \varepsilon \theta_1(s)+\varepsilon^2\theta_2(s)+\varepsilon^3\theta_3(s)+o(\epsilon^3),\\
z(s) &= s + \varepsilon z_1(s)+\varepsilon^2 z_2(s)+\varepsilon^3 z_3(s)+o(\epsilon^3),\\
\alpha(s) &= \varepsilon \alpha_1(s)+\varepsilon^2 \alpha_2(s)+\varepsilon^3 \alpha_3(s)+o(\epsilon^3),
\end{aligned}
\right.\qquad\text{as }\epsilon\rightarrow 0, 
\]
and following the same procedure that was exposed in Section~\ref{sec:small_sigma} for the first and second order terms, we get
\[
\left\{
\begin{aligned}
&\rho_3(s) = -\frac{R_0^2}{8} \left(\cot ^2\left(\frac{\pi }{n}\right)+2\right) \tilde{\sigma}(s)^2 \tilde{\sigma}'(s),\\
&\theta_3(s) = \frac{1}{8 R_0} \cot \left(\frac{\pi }{n}\right)\int_0^s \tilde{\sigma}(\tau ) \left(R_0^2\tilde{\sigma}'(\tau )^2-\csc ^2\left(\frac{\pi }{n}\right) \tilde{\sigma}(\tau )^2\right)\,d\tau,\\
&z_3(s) = \frac{R_0}{24} \cot ^2\left(\frac{\pi }{n}\right) \left(\tilde{\sigma}(s)^3-\tilde{\sigma}(0)^3\right),\\
&\alpha_3(s) = -\frac{R_0}{8} \cot \left(\frac{\pi }{n}\right) \tilde{\sigma}(s)^2 \tilde{\sigma}'(s).\\
\end{aligned}
\right.
\]
\bibliographystyle{abbrvnat}
\bibliography{refs}

\begin{thebibliography}{25}
\providecommand{\natexlab}[1]{#1}
\providecommand{\url}[1]{\texttt{#1}}
\expandafter\ifx\csname urlstyle\endcsname\relax
  \providecommand{\doi}[1]{doi: #1}\else
  \providecommand{\doi}{doi: \begingroup \urlstyle{rm}\Url}\fi

\bibitem[Antman(2005)]{antman_book}
S.~Antman.
\newblock \emph{{Nonlinear Problems of Elasticity}}.
\newblock Springer, 2005.

\bibitem[Arroyo and DeSimone(2014)]{Arroyo_2014}
M.~Arroyo and A.~DeSimone.
\newblock Shape control of active surfaces inspired by the movement of
  euglenids.
\newblock \emph{Journal of the Mechanics and Physics of Solids}, 62:\penalty0
  99--112, 2014.

\bibitem[Arroyo et~al.(2012)Arroyo, Heltai, Millan, and DeSimone]{Arroyo2012}
M.~Arroyo, L.~Heltai, D.~Millan, and A.~DeSimone.
\newblock {Reverse engineering the euglenoid movement}.
\newblock \emph{Proceedings of the National Academy of Sciences}, 109\penalty0
  (44):\penalty0 17874--17879, 2012.
\newblock ISSN 0027-8424.

\bibitem[Audoly and Pomeau(2010)]{audoly2010elasticity}
B.~Audoly and Y.~Pomeau.
\newblock \emph{Elasticity and Geometry: From Hair Curls to the Non-linear
  Response of Shells}.
\newblock Oxford University Press, 2010.

\bibitem[Ba{\v{z}}ant and Cedolin(2010)]{Ba_ant_2010}
Z.~P. Ba{\v{z}}ant and L.~Cedolin.
\newblock \emph{Stability of Structures}.
\newblock World Scientific, 2010.

\bibitem[Bertoldi et~al.(2017)Bertoldi, Vitelli, Christensen, and van
  Hecke]{Bertoldi_2017}
K.~Bertoldi, V.~Vitelli, J.~Christensen, and M.~van Hecke.
\newblock Flexible mechanical metamaterials.
\newblock \emph{Nature Reviews Materials}, 2\penalty0 (11), 2017.

\bibitem[Bigoni et~al.(2015)Bigoni, Dal~Corso, Bosi, and
  Misseroni]{Bigoni_2015}
D.~Bigoni, F.~Dal~Corso, F.~Bosi, and D.~Misseroni.
\newblock Eshelby-like forces acting on elastic structures: Theoretical and
  experimental proof.
\newblock \emph{Mechanics of Materials}, 80:\penalty0 368--374, 2015.

\bibitem[Buetow(1968)]{buetow1968biology}
D.~E. Buetow.
\newblock \emph{The Biology of Euglena: Physiology}.
\newblock The Biology of Euglena. Academic Press, 1968.

\bibitem[Cicconofri and DeSimone(2015)]{Cicconofri_2015}
G.~Cicconofri and A.~DeSimone.
\newblock A study of snake-like locomotion through the analysis of a flexible
  robot model.
\newblock \emph{Proceedings of the Royal Society A: Mathematical, Physical and
  Engineering Sciences}, 471\penalty0 (2184):\penalty0 20150054, 2015.

\bibitem[Cicconofri et~al.(2020)Cicconofri, Arroyo, Noselli, and
  DeSimone]{Cicconofri_2020}
G.~Cicconofri, M.~Arroyo, G.~Noselli, and A.~DeSimone.
\newblock Morphable structures from unicellular organisms with active,
  shape-shifting envelopes: Variations on a theme by {G}auss.
\newblock \emph{International Journal of Non-Linear Mechanics}, 118:\penalty0
  103278, 2020.

\bibitem[Farrell et~al.(2015)Farrell, Birkisson, and Funke]{Farrell_2015}
P.~E. Farrell, {\'{A}}.~Birkisson, and S.~W. Funke.
\newblock Deflation techniques for finding distinct solutions of nonlinear
  partial differential equations.
\newblock \emph{{SIAM} Journal on Scientific Computing}, 37\penalty0
  (4):\penalty0 A2026--A2045, 2015.

\bibitem[Farrell et~al.(2016)Farrell, Beentjes, and
  Birkisson]{farrell2016computation}
P.~E. Farrell, C.~H. Beentjes, and {\'A}.~Birkisson.
\newblock The computation of disconnected bifurcation diagrams.
\newblock \emph{arXiv preprint arXiv:1603.00809}, 2016.

\bibitem[Frenzel et~al.(2017)Frenzel, Kadic, and Wegener]{frenzel2017three}
T.~Frenzel, M.~Kadic, and M.~Wegener.
\newblock Three-dimensional mechanical metamaterials with a twist.
\newblock \emph{Science}, 358\penalty0 (6366):\penalty0 1072--1074, 2017.

\bibitem[Klein et~al.(2007)Klein, Efrati, and Sharon]{Sharon}
Y.~Klein, E.~Efrati, and E.~Sharon.
\newblock Shaping of elastic sheets by prescription of non-{E}uclidean metrics.
\newblock \emph{Science}, 315:\penalty0 1116--1120, 2007.

\bibitem[Leander et~al.(2017)Leander, Lax, Karnkowska, and
  Simpson]{handbook_protists}
B.~S. Leander, G.~Lax, A.~Karnkowska, and A.~G.~B. Simpson.
\newblock \emph{Handbook of the Protists}, chapter Euglenida.
\newblock Springer, 2017.

\bibitem[Lessinnes and Goriely(2016)]{Lessinnes_2016}
T.~Lessinnes and A.~Goriely.
\newblock Design and stability of a family of deployable structures.
\newblock \emph{{SIAM} Journal on Applied Mathematics}, 76\penalty0
  (5):\penalty0 1920--1941, 2016.

\bibitem[Logg et~al.(2012)Logg, Mardal, Wells, et~al.]{LoggMardalEtAl2012a}
A.~Logg, K.-A. Mardal, G.~N. Wells, et~al.
\newblock \emph{Automated Solution of Differential Equations by the Finite
  Element Method}.
\newblock Springer, 2012.
\newblock ISBN 978-3-642-23098-1.

\bibitem[Moakher and Maddocks(2005)]{Moakher_2005}
M.~Moakher and J.~H. Maddocks.
\newblock A double-strand elastic rod theory.
\newblock \emph{Archive for Rational Mechanics and Analysis}, 177\penalty0
  (1):\penalty0 53--91, 2005.

\bibitem[Neukirch and Van Der~Heijden(2002)]{neukirch2002geometry}
S.~Neukirch and G.~H.~M. Van Der~Heijden.
\newblock Geometry and mechanics of uniform n-plies: from engineering ropes to
  biological filaments.
\newblock \emph{Journal of Elasticity}, 69\penalty0 (1-3):\penalty0 41--72,
  2002.

\bibitem[Noselli et~al.(2019{\natexlab{a}})Noselli, Arroyo, and
  DeSimone]{Noselli_2019}
G.~Noselli, M.~Arroyo, and A.~DeSimone.
\newblock Smart helical structures inspired by the pellicle of euglenids.
\newblock \emph{Journal of the Mechanics and Physics of Solids}, 123:\penalty0
  234--246, 2019{\natexlab{a}}.

\bibitem[Noselli et~al.(2019{\natexlab{b}})Noselli, Beran, Arroyo, and
  DeSimone]{Noselli2019a}
G.~Noselli, A.~Beran, M.~Arroyo, and A.~DeSimone.
\newblock {Swimming Euglena respond to confinement with a behavioural change
  enabling effective crawling}.
\newblock \emph{Nature Physics}, 15\penalty0 (5):\penalty0 496--502,
  2019{\natexlab{b}}.
\newblock ISSN 1745-2473.

\bibitem[Starostin and van~der Heijden(2014)]{Starostin_2014}
E.~Starostin and G.~van~der Heijden.
\newblock Theory of equilibria of elastic 2-braids with interstrand
  interaction.
\newblock \emph{Journal of the Mechanics and Physics of Solids}, 64:\penalty0
  83--132, 2014.

\bibitem[Thompson et~al.(2002)Thompson, van~der Heijden, and
  Neukirch]{Thompson_2002}
J.~M.~T. Thompson, G.~H.~M. van~der Heijden, and S.~Neukirch.
\newblock Supercoiling of {DNA} plasmids: mechanics of the generalized ply.
\newblock \emph{Proceedings of the Royal Society of London. Series A:
  Mathematical, Physical and Engineering Sciences}, 458\penalty0
  (2020):\penalty0 959--985, 2002.

\bibitem[Tondu(2012)]{Tondu_2012}
B.~Tondu.
\newblock Modelling of the {McKibben} artificial muscle: A review.
\newblock \emph{Journal of Intelligent Material Systems and Structures},
  23\penalty0 (3):\penalty0 225--253, 2012.

\bibitem[Zhao et~al.(2014)Zhao, Zhao, Wang, Zhang, and Feng]{Zhao_2014}
Z.-L. Zhao, H.-P. Zhao, J.-S. Wang, Z.~Zhang, and X.-Q. Feng.
\newblock Mechanical properties of carbon nanotube ropes with hierarchical
  helical structures.
\newblock \emph{Journal of the Mechanics and Physics of Solids}, 71:\penalty0
  64--83, 2014.

\end{thebibliography}
\end{document}